\documentclass{acm_proc_article-sp}

\usepackage{graphicx}
\usepackage{color}
\usepackage{subfigure}

\begin{document}

\newcommand{\xhdr}[1]{\paragraph*{\bf {#1}}}

\title{Assembling thefacebook: Using Heterogeneity to Understand Online Social Network Assembly}

\numberofauthors{4} 
\author{
\alignauthor
Abigail Z.\ Jacobs\\
       \affaddr{University of Colorado Boulder}\\
       \affaddr{abigail.jacobs@colorado.edu}
\alignauthor
Samuel F.\ Way\\
       \affaddr{University of Colorado Boulder}\\
       \affaddr{samuel.way@colorado.edu}
\alignauthor 
Johan Ugander\\
       \affaddr{Microsoft Research}\\
       \affaddr{Stanford University}\\
       \affaddr{jugander@stanford.edu}
\and 
\alignauthor Aaron Clauset\\
       \affaddr{University of Colorado Boulder}\\
       \affaddr{Santa Fe Institute}\\
       \affaddr{aaron.clauset@colorado.edu}
}

\maketitle
\begin{abstract}
Online social networks represent a popular and diverse class of social media systems. Despite this variety, each of these systems undergoes a general process of \textit{online social network assembly}, which represents the complicated and heterogeneous changes that transform newly born systems into mature platforms. However, little is known about this process. For example, how much of a network's assembly is driven by simple growth? How does a network's structure change as it matures? How does network structure vary with adoption rates and user heterogeneity, and do these properties play different roles at different points in the assembly? We investigate these and other questions using a unique dataset of online connections among the roughly one million users at the first 100 colleges admitted to Facebook, captured just 20 months after its launch. We first show that different vintages and adoption rates across this population of networks reveal temporal dynamics of the assembly process, and that assembly is only loosely related to network growth. We then exploit natural experiments embedded in this dataset and complementary data obtained via Internet archaeology to show that different subnetworks matured at different rates toward similar end states. These results shed light on the processes and patterns of online social network assembly, and may facilitate more effective design for online social systems.
\end{abstract}

\section{Introduction\label{sec:intro}} 

Since their emergence in the mid-1990s, online social networks have grown into a highly popular and diverse class of social media systems. This class includes now-defunct systems such as Friendster, tribe.net and Orkut, niche systems such as Academia.edu and HR.com, and large, more general systems such as Facebook and LinkedIn. In contrast to earlier online social communities such as newsgroups~\cite{fisher2006you} and weblogs~\cite{marlow2004audience}, many modern systems tend to encourage users to transfer offline relationships onto an online setting. 
Despite the wide variety of these systems---professional vs.\ personal, contextual vs.\ general, virtual vs.\ anchored offline---all of these systems undergo a general process of \textit{online social network assembly} that represents the complicated and heterogeneous changes by which newly born systems evolve into mature platforms.

Relatively little, however, is known about the central tendencies or variability of this process, while such understanding would shed considerable light on the effective design of new platforms. As a result, questions abound. How much of a network's assembly is driven by simple growth processes? How does a network's structure change as it matures? How does network structure vary with adoption rates and user heterogeneity, and do these properties play different roles at different points in the assembly? Are there distinct developmental ``phases'' to the assembly of these systems?

One reason we lack good answers to such questions is a lack of good data. Traditional online social network datasets fall short in two key ways. First, understanding the effects of different processes requires a network-population perspective, in which many parallel network instances can be examined in order to discern the natural variability of network structure. Second, in the rare situations where populations of networks have been available, such as the National Longitudinal Study of Adolescent Health~\cite{resnick1997protecting}, the underlying social processes do not vary across network instances enough to identify and model different aspects of assembly.  By analogy, in social networks recorded from survey questionnaires, it is well-known that different so-called {\it name generators}~\cite{campbell1991name}---questions used to elicit social ties---lead to networks with substantially different structure. As a broad generalization for online social networks, we are interested in the general consequences of variations in the circumstances under which social networks are assembled online. 

To understand the structural impact of different assembly processes, we therefore need a population of networks that vary dependably in their assembly. The so-called \textit{Facebook100} dataset~\cite{traud2012social}, which is a snapshot of 100 within-college social networks on Facebook in September 2005, provides just such a population. These networks provide a unique perspective on the very early assembly of a major online social network platform. Crucial to our investigation, these networks vary somewhat in their sizes, characteristics, and history. Each network has a different ``vintage,'' representing a different amount of time between when the college first adopted Facebook and when the snapshot was taken. These vintages, and differential adoption rates across colleges, effectively reveal temporal dynamics of the assembly processes, which we exploit. Finally, a series of natural experiments related to the academic calendar and college characteristics created sufficient heterogeneity at the user- and network-level, which in turn can reveal certain aspects of the underlying assembly processes.

\begin{figure}[t!]
\begin{center}
\includegraphics[width=0.485\textwidth]{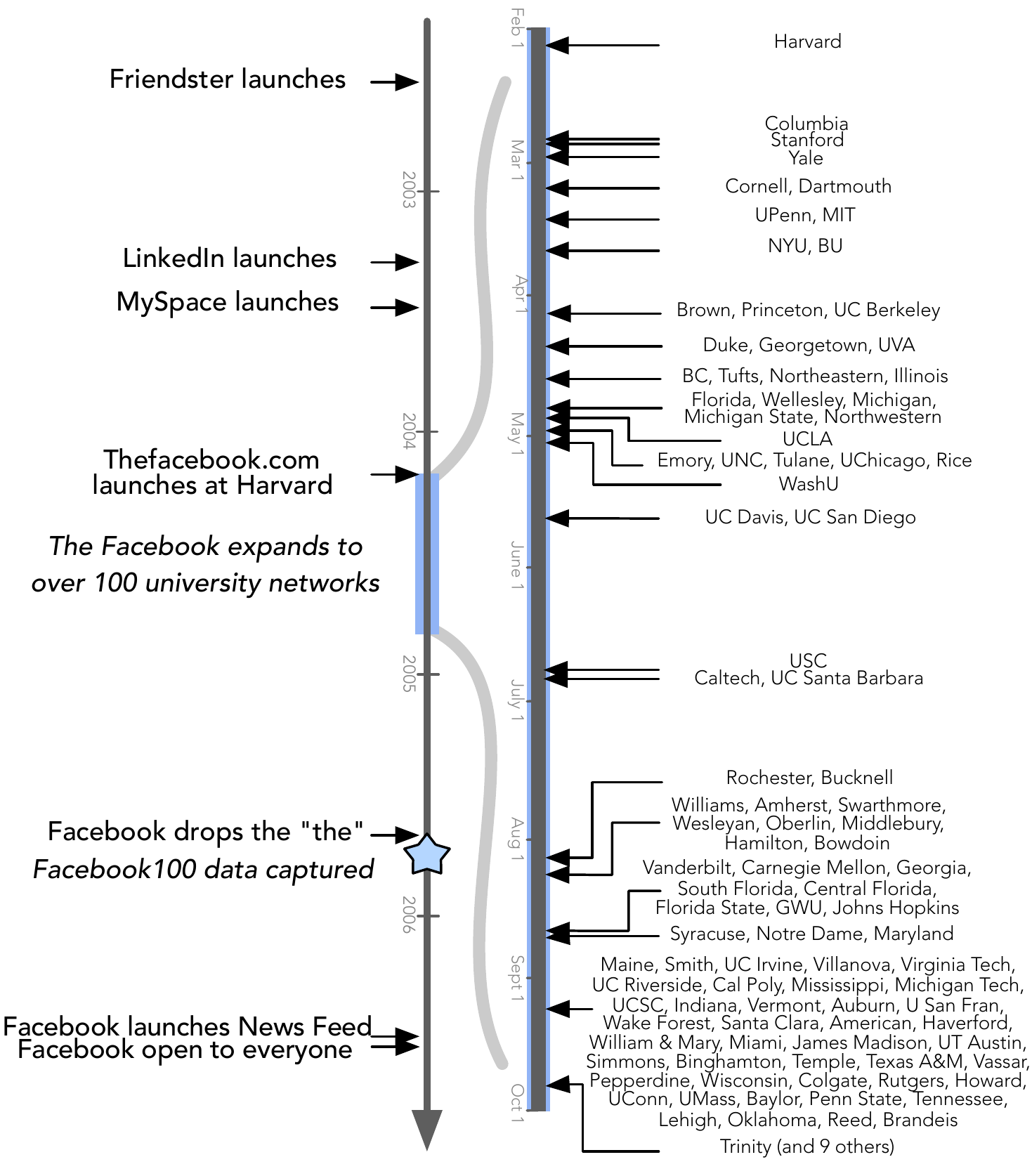} 
\caption{Key milestones in the early history of Facebook, including launch dates for the 100 colleges in the Facebook100 dataset.
\newline
}
\label{fig:FBtimeline}
\end{center}
\end{figure}

As an example of a natural experiment we can exploit, we note that these 100 colleges joined Facebook sometime between its launch in February 2004 and the end of September 2004 (Fig.~\ref{fig:FBtimeline}). Because this period spans the end of the 2003--2004 school year, students in some graduating classes of 2004 would have experienced Facebook only as alumni (colleges that joined after graduation) while others experienced it as students (colleges that joined earlier). Comparing the subnetworks of these two groups of students, who should otherwise be fairly similar, with each other and with students of earlier or later graduation years, will shed light on the importance of physical proximity and on-campus interactions in driving network assembly.

As an additional natural experiment, the networks were observed in early September 2005, during the beginnings of the 2005--2006 academic calendars, dates that again vary considerably in this population. As a result, students in the class of 2009 (incoming freshmen in 2005) enrolling at colleges with late start dates (late September) were observed before any significant offline interactions could have taken place (excluding brief summer orientation programs and students from the same high schools). As the students in these classes largely lack any shared historical context, the networks corresponding to colleges with late start dates primarily represent assemblies of relationships formed online, rather than offline. In contrast, students in the class of 2009 enrolling at colleges with early start dates \textit{will} have shared a real world context. This affords an opportunity to ask: how do online social networks encoding online interactions differ in structure from networks that are also encoding offline interactions?  We address this by constrasting the classes of 2009 at these early- and late-starting colleges.

By complementing the Facebook100 dataset with the above dates (a modest Internet-archaeological effort\footnote{These data supplements are available at \texttt{http://azjacobs.com/fb100/} and in the appendix.}), as well as with basic statistics provided by the U.S. Department of Education, we provide a unique, discerning perspective into how online social network structures differ depending on (i) the presence or absence of an underlying offline social network (by studying the classes of 2009), and (ii) the presence or absence of present-time social interactions (by studying the classes of 2004). We also present broad analyses of the population-level variability of network statistics in a general assembly process observed at different vintages. These results shed new light on the general processes that shape social network assembly in online environments, and may facilitate more effective designs of online social systems that relate to the offline world. 

\section{Facebook in the age of \newline Friendster}\label{sec:FBageFriendster}

In 2015, Facebook is today a large and sophisticated social media system, claiming more than \mbox{1.44 billion} monthly active users (as of March 2015). In 2005, however, Facebook was a very different kind of online social network, in a correspondingly different social media landscape~\cite{boyd2007SNSdefinition}. 

\begin{figure}[t!]
\begin{center}
\subfigure{\includegraphics[width=0.225\textwidth]{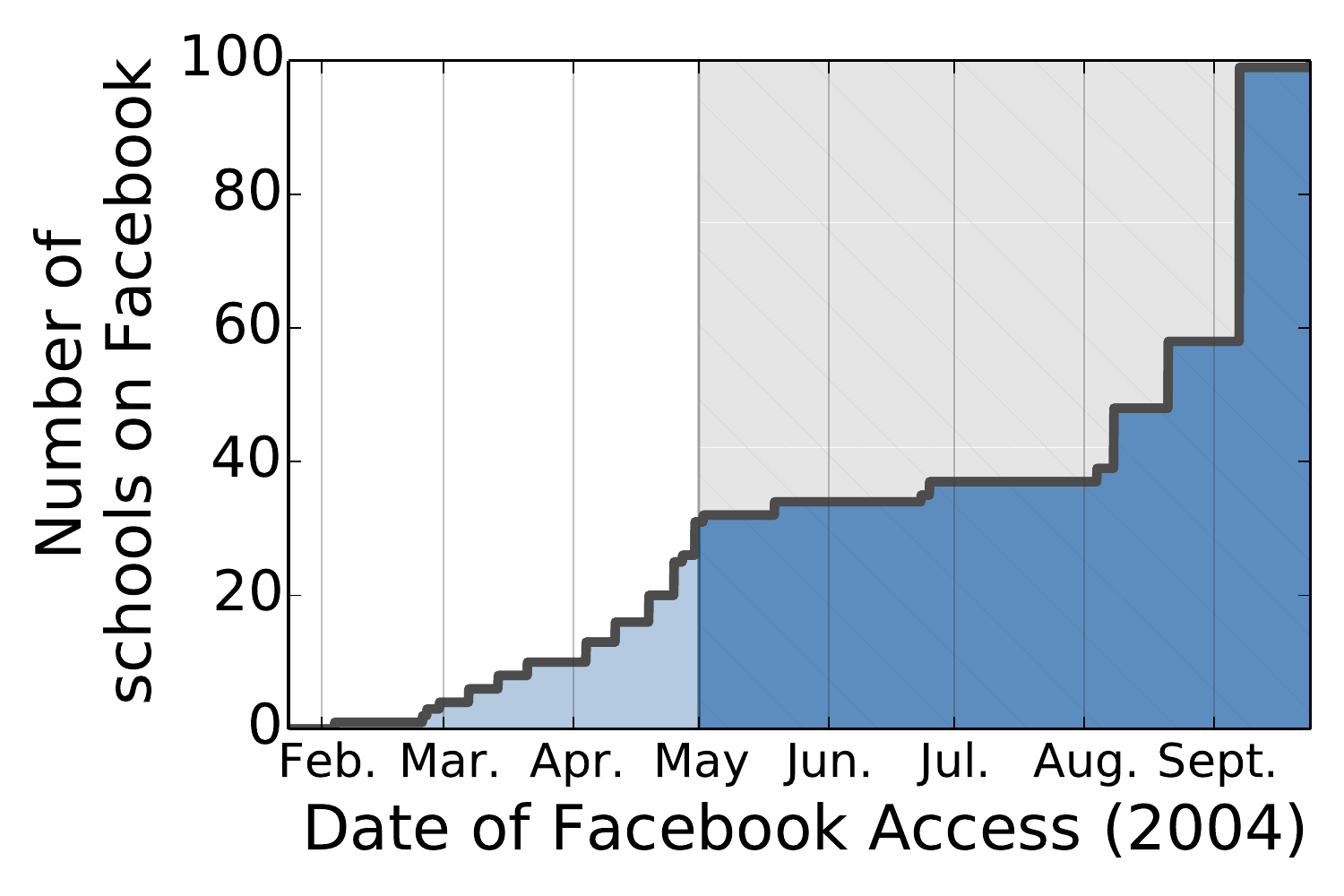}}
\subfigure{\includegraphics[width=0.225\textwidth]{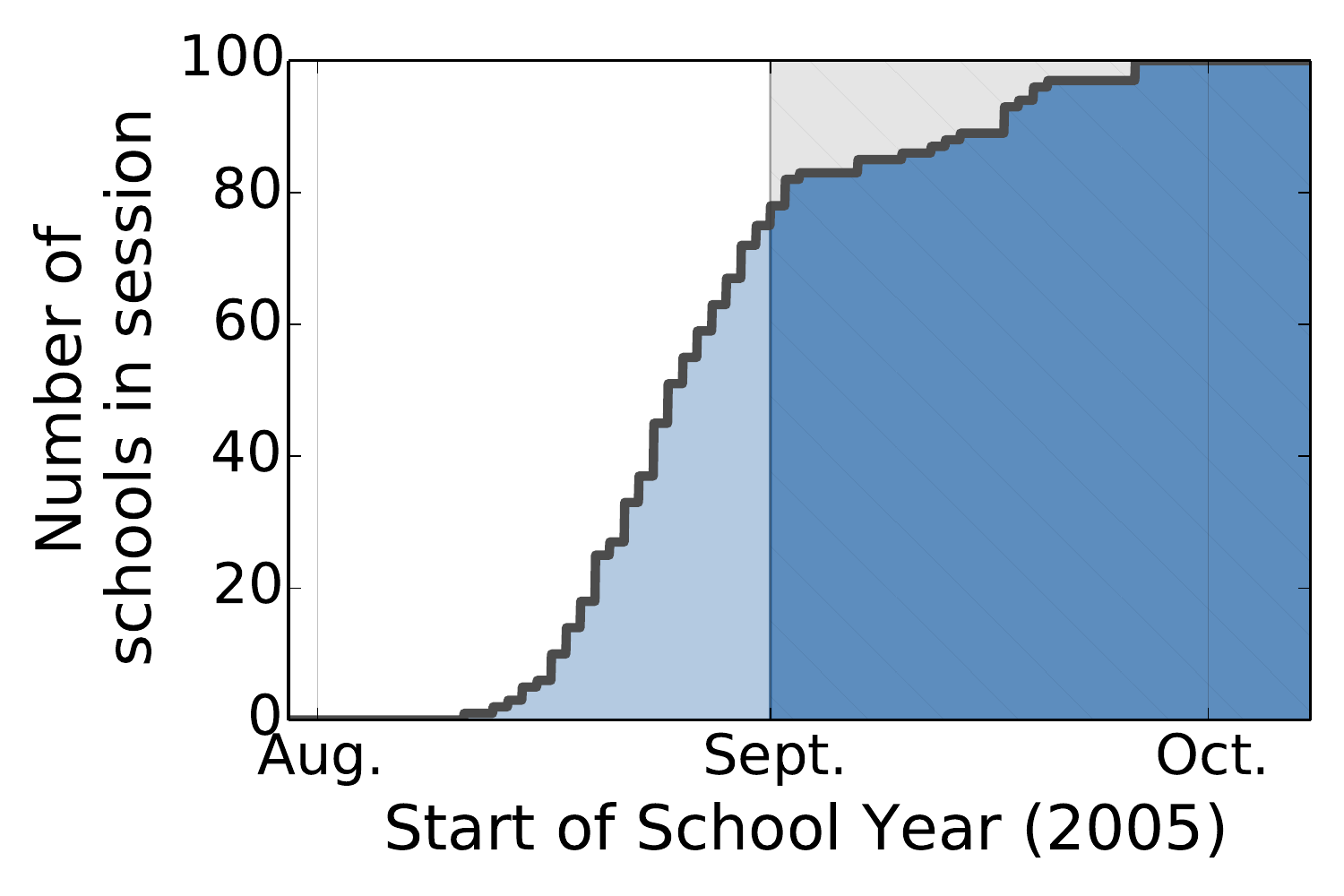}}
\caption{The cumulative distribution of schools in the Facebook100 dataset, by date added to Facebook during 2004 (left) and by start of the 2005--2006 school year (right). Shaded regions show how colleges are divided in terms of having received access to Facebook before or after the end of the 2003--2004 school year and whether or not the 2005--2006 school year had begun when the Facebook100 dataset was collected.}
\label{fig:cdfs}
\end{center}
\end{figure}

Facebook launched at Harvard University on February 4th, 2004 under the name \texttt{thefacebook.com}, at a time when the dominant online social networks were Friendster and MySpace. A host of other online college ``facebooks'' such as CUcommunity, CampusNetwork, and CollegeFacebook were also emerging, in addition to efforts by individual universities to move their student directories onto the Web. Facebook initially limited registration to users affiliated with a sanctioned but growing list of colleges, starting with Harvard (Figs.~\ref{fig:FBtimeline} and~\ref{fig:cdfs}). Facebook's popularity spread quickly\footnote{ \textit{The Daily Northwestern} describes the first 48 hours of Facebook access at Northwestern University thusly: `` `It's an epidemic$\ldots$ my whole hall is infected,' said Erica Birnbaum, a Communication freshman. But it's not only one hall.  After being available for only about 34 hours, 931 NU students already had registered as of 8 p.m.\ Monday $\ldots$ Such a large quantity of friend request and confirmation e-mails being sent from the Facebook caused Northwestern University Information Technology to block all mail sent from the site Sunday night$\ldots$ `It was viewed as an attack against the network.' '' (26 April 2004)}, and by the time of the Facebook100 snapshot (September 2005), Facebook had dropped the ``the'' in its name, opened to over 800 colleges, and had just begun opening itself to high school students.
By December 2005, Facebook's user base numbered 6 million, compared to 20 million for Friendster and over 22 million users for MySpace. In September 2006, Facebook opened to all persons over the age of 13.

\xhdr{Description of the network dataset}
The Facebook100 dataset~\cite{traud2012social} contains an anonymized snapshot of the friendship connections among  \mbox{$n=1,208,316$} users affiliated with the first 100 colleges admitted to Facebook, all located in the United States. This comprises a total of \mbox{$m=93,969,074$} friendship edges (unweighted and undirected) between users within each separate college. Each vertex is associated with an array of social variables representing the person's status (undergraduate, graduate student, summer student, faculty, staff, or alumni), dorm (if any), major (if any), gender~(M~or~F), and graduation year. Across all networks, only 0.03\% of status values are missing. Other variables have slightly higher missing rates (gender: 5.6\%; graduation year: 9.8\%). Dorm and major have higher rates still, which is likely related to off-campus living and undeclared majors. The completeness of these data reflects the pervading social norms surrounding data privacy expectations in 2005, and possibly a selective bias against users who disliked the default setting of sharing all information within the college network~\cite{acquisti2006imagined,tufekci2008grooming}.

For nearly all colleges, alumni made up about 10--25\% of users, a quantity that increased with the age of the network. Vertices labeled as faculty, staff or students who were not regular undergraduates (graduate students and summer students) made up on average 4.1\% of each population.

Each college network includes an ``index'' variable that gives its ordinal position of when it joined Facebook: Harvard is 1 and Trinity College is 100 (Fig.~\ref{fig:FBtimeline}). For each network, we acquired college-level variables (enrollment, public vs.\ private, semester vs.\ quarter calendar) from the Integrated Postsecondary Education Data System (IPEDS) provided by the U.S. Department of Education~\cite{ipeds}. Full-time undergraduate enrollment from 2007, the earliest date for which data are fully available, was used a proxy for 2005 enrollment.

\begin{figure}[t!]
\begin{center}
\includegraphics[width=0.48\textwidth]{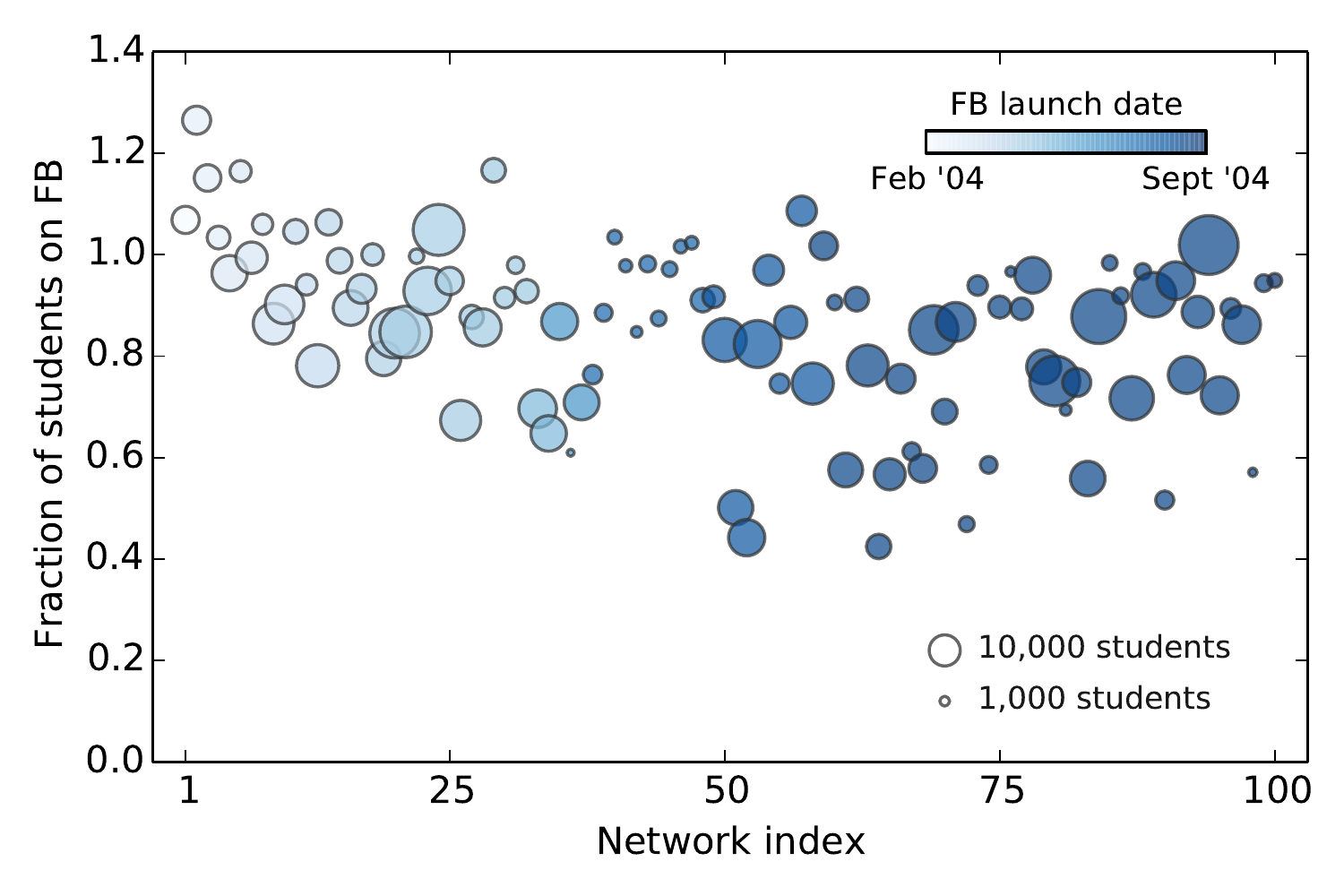}
\caption{
Fraction of undergraduates that adopted Facebook vs.\ network index. 
Vintage is visualized with network index, the order in which schools were given access to the site. Size corresponds to the size of the undergraduate population. Color indicates the date on which schools were opened to Facebook. 
}
\label{fig:sizevsage}
\end{center}
\end{figure}

By dividing the number of undergraduate accounts in each college network by reported enrollment, we can estimate the fraction of students in each network who were on Facebook, a measure of service adoption (Fig.~\ref{fig:sizevsage}). In some cases, the estimated ratio exceeds 1.0 as a result of either errors in our enrollment numbers, part-time students on Facebook who were not counted as ``full-time enrolled,'' or multiple/fake accounts at the few colleges that allowed students to control multiple email aliases and circumvent Facebook's initial limits on access.

\section{Online social network assembly}
Online social network assembly is the process by which networks transform from initial creation to a mature online social network. Assembly processes are affected by the composition of the community, online and offline social and behavioral practices, limits on growth (e.g., needing an elite university email address), and competition from other systems, among other mechanisms. Assembly can in part be characterized by the sequence of structural changes that newly-born online social networks undergo as they mature. In particular, this area of study aims to identify and model the underlying social processes that guide assembly, and to identify the `developmental' patterns that are common across different networks. Here, we focus on the role of network growth, user heterogeneity, adoption rate, and network `vintage' in shaping these assembly patterns. We examine the impact of these elements on structural patterns in the networks, e.g., their degree distributions, clustering coefficients, diameters, and community structure~\cite{ugander2011anatomy}, as well as understanding how those patterns change under network growth~\cite{backstrom2012four}, how they vary across subpopulations within the network, and what social processes govern these patterns and variations. We note, however, that reliably connecting observed patterns with the correct underlying processes can be complicated, as different processes can sometimes lead to similar, or even identical, structural patterns~\cite{mitzenmacher2004powerlaw}.

By comparing patterns across these networks, we aim to characterize the scale and sources of natural variation. Here, several observable features of Facebook's early college networks---differing potential network sizes, ages, adoption and heterogeneity of context---play important roles in shedding light on its early assembly. First, its staged expansion among colleges during 2004 produced a population of online social networks of different vintages, at schools of different sizes, within which the service was adopted at different rates. Second, the graduation year annotations identify subpopulations that changed identity during the time observed, e.g., different classes that joined or left the campus environment.

\xhdr{Processes and models of assembly}
Online social network assembly is a special kind of network evolution. Most techniques and statistical models developed for analyzing the structure of temporal networks~\cite{holme2012temporal}, however, cannot be applied to the Facebook100 data because these networks are not snapshots of a single evolving system. Instead, we will exploit the several ways that temporal information is embedded within the observed network structures and represented in their covariates, e.g., vintage and adoption rates at the network level and graduation year at the vertex level.

The simplest model of assembly is network growth, in which the number of vertices and edges grow monotonically in some way. Several simple models of network growth exist, including many variations on preferential attachment~\cite{kumar2010structure}, in which new users join the network and create connections with existing users with probability proportional to those users' current degree; randomly grown networks~\cite{callaway2001randomly}, which are related to classic random graph models; and the forest-fire model~\cite{leskovec2005densificationKDD}, which is related to preferential attachment but produces both greater local clustering and a shrinking diameter. Crucially, these models assume that assembly is a homogeneous process, and thus network structure changes uniformly across all subsets of vertices~\cite{schoenebeck2013potential}. In contrast, the assembly patterns of real online social networks are likely to be considerably more heterogeneous, both at the vertex level and at the network level. These models thus hold value primarily as theoretical reference points in our analysis.

Social surveys of early Facebook users provides some hints about the processes governing its assembly, and support our claim that assembly in real networks is unlikely to be simple or homogeneous. One survey from 2006 found that students of different graduating years had different usage patterns, and that older students---those whose college careers were mostly over by the time Facebook arrived on their campus---were less likely to adopt the service~\cite{tufekci2008grooming}. Thus, local network structure is likely to vary by graduating year. Several surveys also found evidence that online connections on Facebook among current students generally reflect pre-existing offline relationships~\cite{lampe2006face,mayer2008old}. This implies that Facebook's early assembly should reflect the inhomogeneities of real-world social processes, which depend on factors like age, gender, and being on campus.

From a theoretical perspective, the social processes that seem likely to influence assembly in these networks can be divided into two major dichotomies:\ {offline/online processes} and {contemporary/historical processes}. In the first case, offline processes are those driven by relationships in the offline world that are then transferred to an online setting, while online processes are confined to mechanisms mediated by digital interactions alone. In the second case, contemporary processes are those that reflect social events in the present time, while historical processes are those where the formation of links in the online social network is driven by pre-existing relationships that are brought online.

These classes represent different ways that social connections can be recorded in online networks, and are orthogonal to the social processes that drive link formation, such as homophily, social status, or strategic behavior~\cite{mcpherson2001birds, Ball2012, burt2009structural}. For instance, triadic closure---the event in which two people who have a mutual friend, but who are not themselves currently friends, become friends---can drive relationships in the past or present, because closing a triad can occur at any time, and can be mediated by either offline or online interactions. Different endogenous or exogenous forces can also shape the assembly of a particular online social network. For instance, features like Facebook's ``People You May Know'' module influence which links form by facilitating the transfer of offline relationships to the online network~\cite{zignani2014triadic}, while competition from other systems can impede or reverse link formation altogether~\cite{ribeiro2014modeling}. The systematic loss of links, and more generally the decay and disassembly of online social networks is a related but distinct research domain, as disassembly processes are not simply assembly processes in reverse~\cite{bascompte2009assembly,garcia2013friendsterautopsy}. 

Here we focus on three distinct types of social processes in our data, and how they relate to the network assembly of early Facebook:\ (i)~the transfer of offline historical friendships to the online environment~\cite{ellison2007benefits}, (ii)~the formation of connections that reflect present day and offline interactions in the college environment, and (iii)~connections formed purely online. We expect to observe a mixture of these processes, and the corresponding patterns they induce, across our network population. Furthermore, because past work suggests that Facebook connections, from the very start, reflected offline social interactions~\cite{lampe2006face,mayer2008old}, we expect that networks further along in the assembly process will more closely resemble complex offline social structures. We expect strong differences in how quickly different Facebook subnetworks assemble, for instance between students and alumni, because students often live together, take classes together, socialize and work together and alumni generally do not. 

Network growth due to accretion, in which existing users invite their friends to the network, and due to triadic closure mechanisms would tend to make the more mature subnetworks appear more dense, with higher mean degrees, and lower mean geodesic distances than less mature subnetworks. We expect the differences between subnetworks to decrease with older vintages. In addition, we expect different subnetworks to mature at different rates, unlike previous work that focuses on homogeneous processes~\cite{schoenebeck2013potential}.

Finally, given Facebook's role in 2005 as a campus-oriented social network, we expect that adoption among undergraduates can be used as a proxy for maturity of the network. As the early design was to facilitate within-campus interactions, the college online social networks would grow by adding new users and increasing the connections among them. High adoption indicates the online social network would be nearing its effective finite limit for the undergraduate network.


\section{Vintage, growth, and adoption in network assembly \label{sec:networkassembly}}

To begin our analysis, we first test how changes in network structure are related to network size, network vintage, and service adoption. While the domain of study about network growth investigates the relationship of network properties to network size, it is an open question whether network assembly can be strictly explained by network size or network vintage, the relationship to which is not obvious \textit{a priori}. We thus expect to see either no relationship between a particular measure of network structure and age---in the case that the corresponding network property is roughly stationary under the assembly process---or a simple relationship---in the case that the property is gradually modified with age. Alternatively, if assembly is equivalent to simple growth, as in traditional network models of growth, we expect to see certain specific relationships between network measures and network size. We evaluate these two competing hypotheses by examining the relationship of standard measures of network structure, such as mean degree, clustering coefficient, mean geodesic distance, and degree assortativity with network size $n$ and vintage.\footnote{For clarity, we visualize the schools by network \textit{index}, corresponding to the order in which schools were added to Facebook. In these cases we overlay the color corresponding to the date added (Fig.~\ref{fig:FBtimeline}), thereby vintage is monotonically increasing with index.}

\begin{figure}[t!]
\begin{center}
\includegraphics[width=0.495\textwidth]{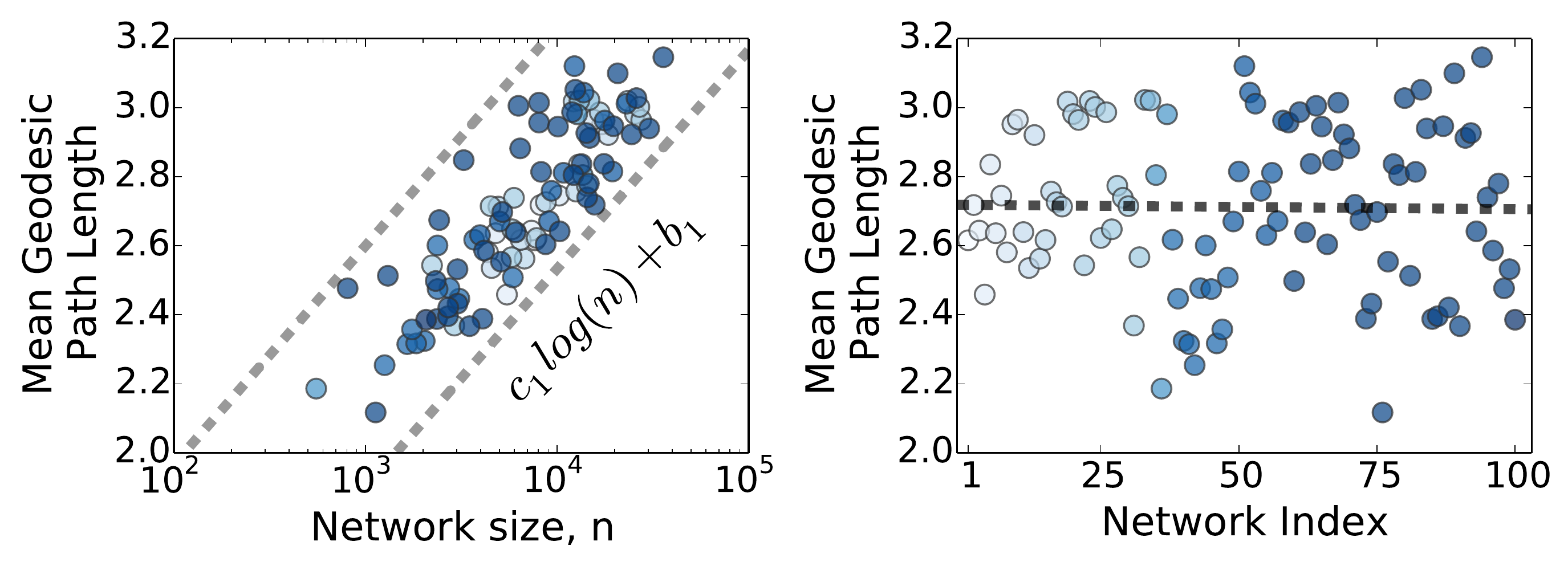}
\includegraphics[width=0.495\textwidth]{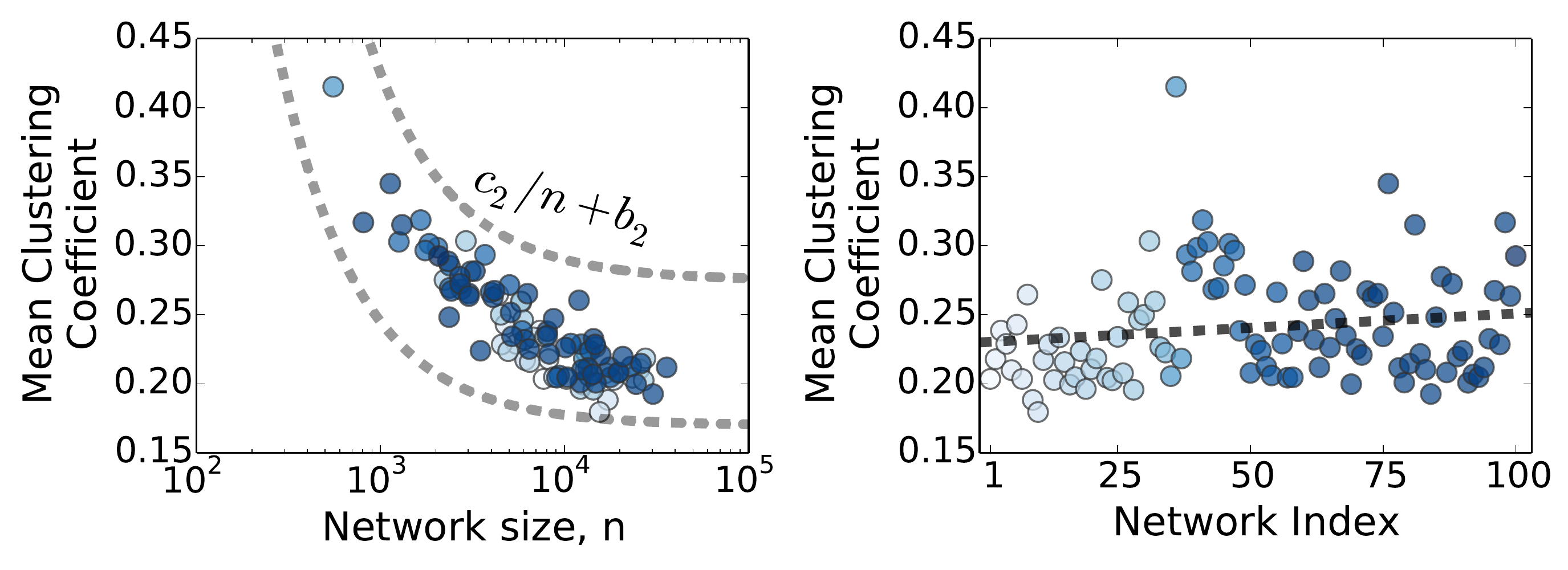}
\caption{(top) Mean geodesic distance (shortest path length), and (bottom) mean clustering coefficient ordered by school size $n$ and by network index. In agreement with results from random graph theory, the mean geodesic distance varies like $O(\log n)$ and the clustering coefficient varies like $1/n$. Color indicates the vintage of the network by date added. Dashed lines show an ordinary least squares fit to the data, demonstrating little to no trend between network features and vintage.
}
\label{fig:geodesicsize}
\end{center}
\end{figure}

We find that these networks as a population exhibit the classic ``small world'' pattern found in many social networks, with small pairwise distances and relatively high average clustering coefficients, capturing the frequency of triangles to length-two paths~\cite{watts1998collective}. Specifically, the mean geodesic distance (average length of a shortest path) scales like $O(\log n)$ with network size $n$, while the clustering coefficient scales like $O(1/n)$, seemingly towards a modest constant as an asymptotic end state (Fig.~\ref{fig:geodesicsize}); in contrast, neither mean geodesic distance nor clustering coefficient varies clearly with vintage. The rising mean geodesic distance with $n$, and its independence of vintage, contrasts with the graph densification literature~\cite{leskovec2005densificationKDD}, which predicts a falling distance with size or time, and it is instead consistent with basic theories for random graphs, which predicts a $O(\log n)$ behavior. The fact that a densification pattern is observed in Facebook several years later~\cite{backstrom2012four} suggests that online social network assembly may go through distinct developmental phases, with an early phase of sparsification, resembling a growing random graph~\cite{callaway2001randomly}, that is followed much later by densification. The falling clustering coefficient pattern observed here, which is expected in random graphs but not in social networks~\cite{newman:networks:book}, supports this hypothesis.

\begin{figure}[t!]
\begin{center}
\includegraphics[width=0.45\textwidth]{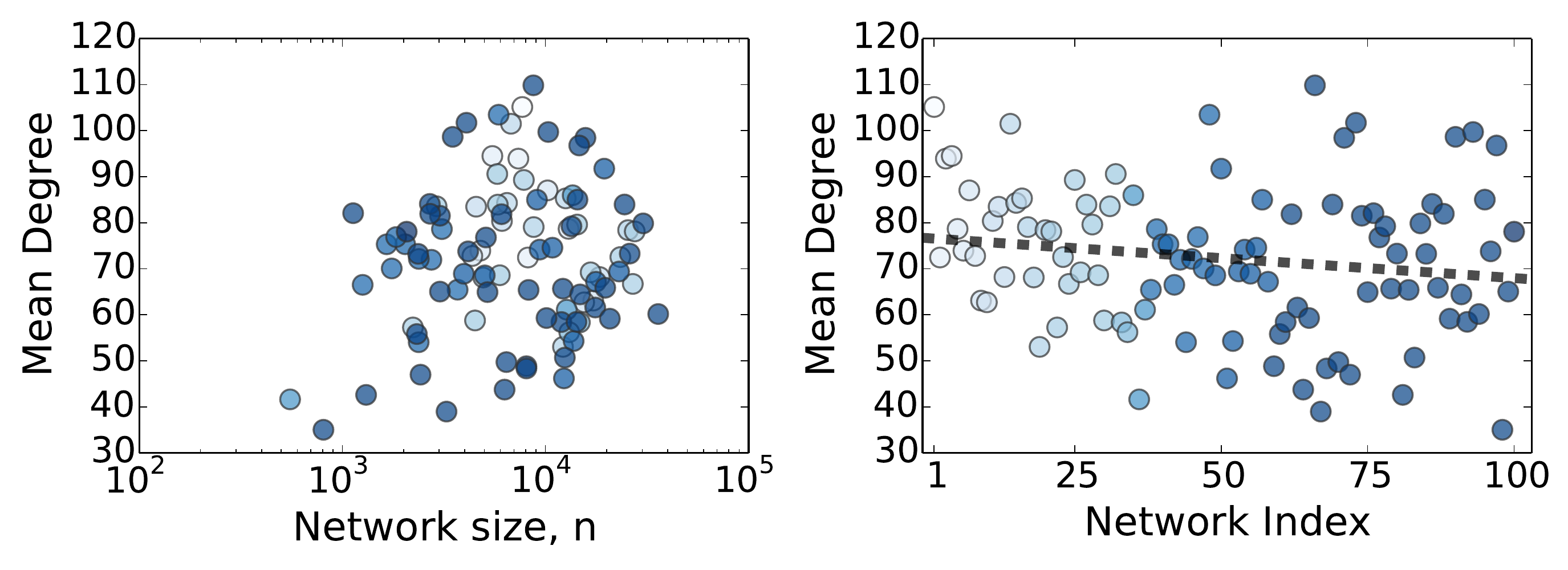}
\includegraphics[width=0.45\textwidth]{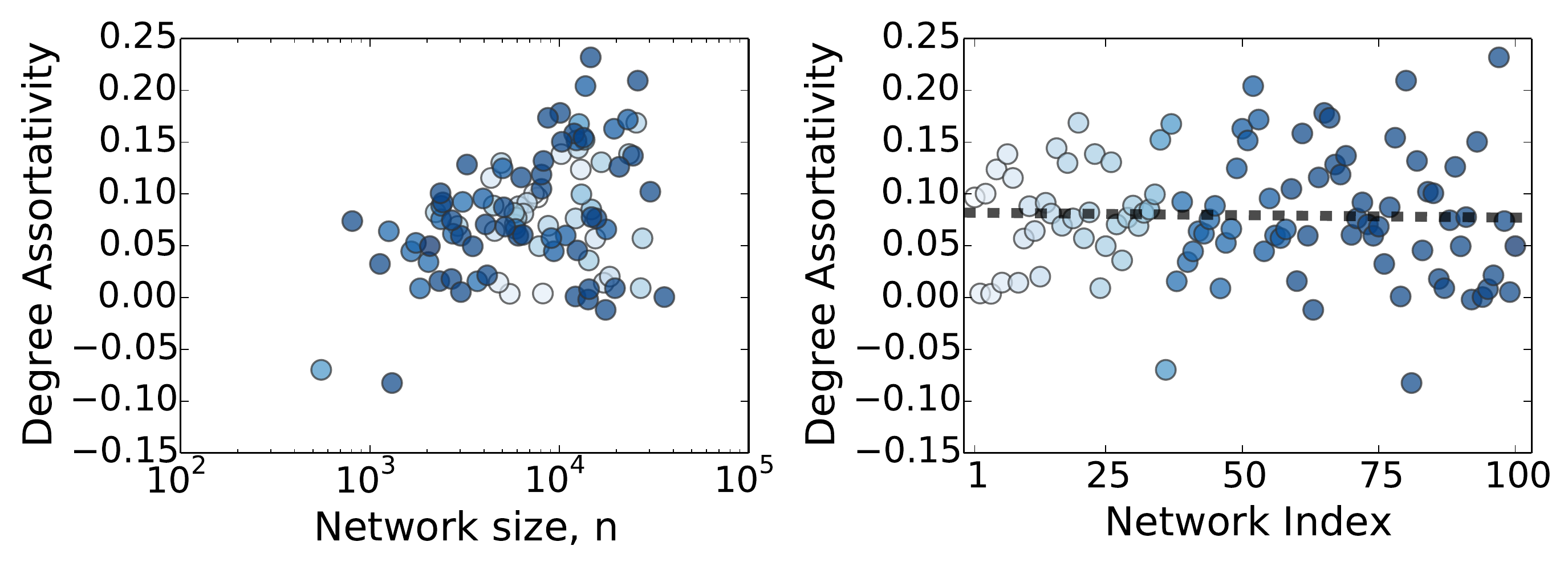}
\includegraphics[width=0.45\textwidth]{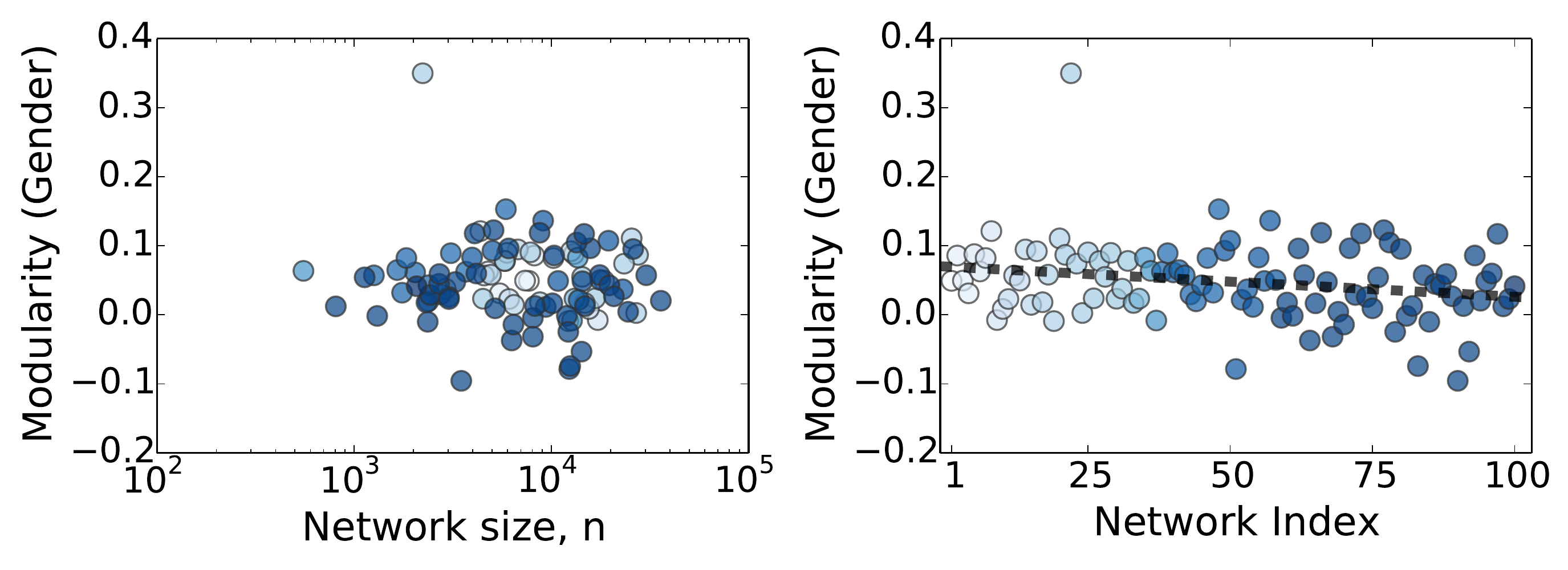} 
\includegraphics[width=0.45\textwidth]{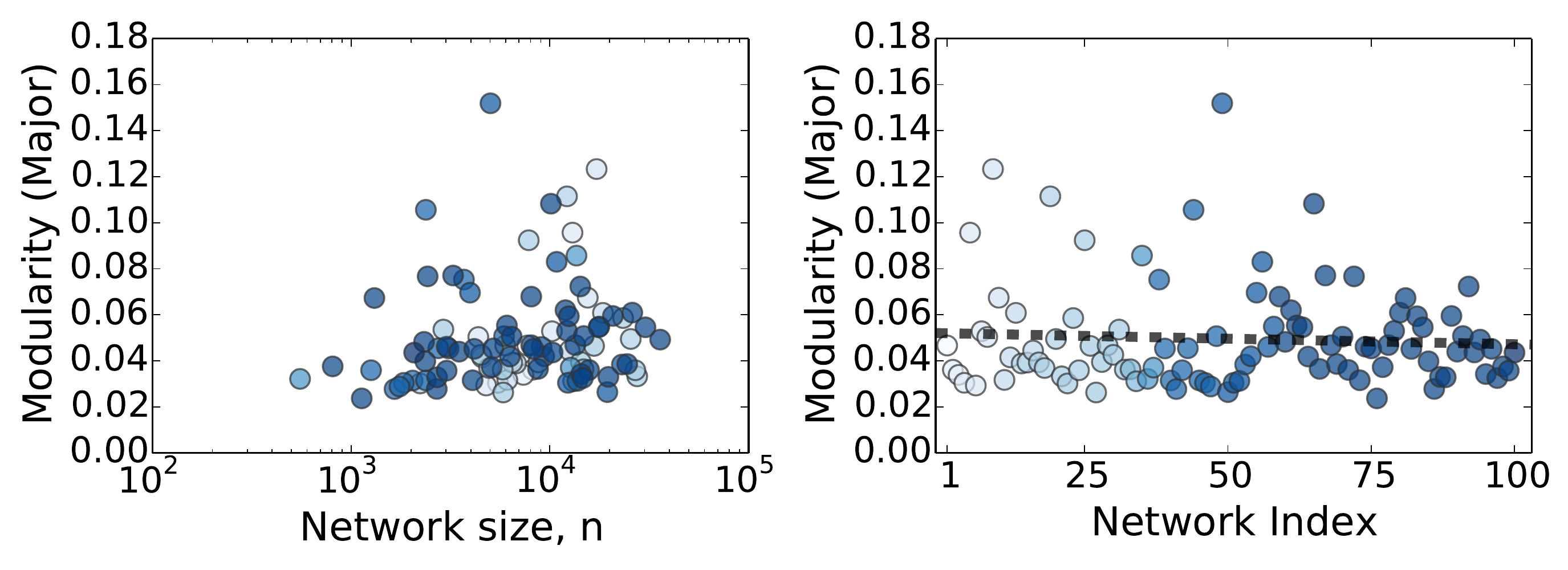} 
\caption{Relation of various network features to network size and network index. Colors indicate the vintage of the network by date added. Dashed lines show an ordinary least squares fit to the data, demonstrating little to no trend between network features and vintage.
}
\label{fig:sizevsagefeatures}
\end{center}
\end{figure}

We examine several other measures of network structure, such as mean degree; assortativity on vertex degree (Pearson correlation of degrees between connected pairs); and modularity by gender or major. Modularity quantifies whether pairs share an attribute more than expected by random (positive) or less (negative)~\cite{newman:networks:book}. We find very weak or no correlation with network size or network vintage (Fig.~\ref{fig:sizevsagefeatures}). The lack of any clear relation with size and vintage for these measures supports the notion that the online social network assembly process for Facebook college networks is not uniquely explained by size and vintage. That is, assembly is more complex than simple growth or network vintage.

\begin{figure}[t!]
\begin{center}
\includegraphics[width=0.30\textwidth]{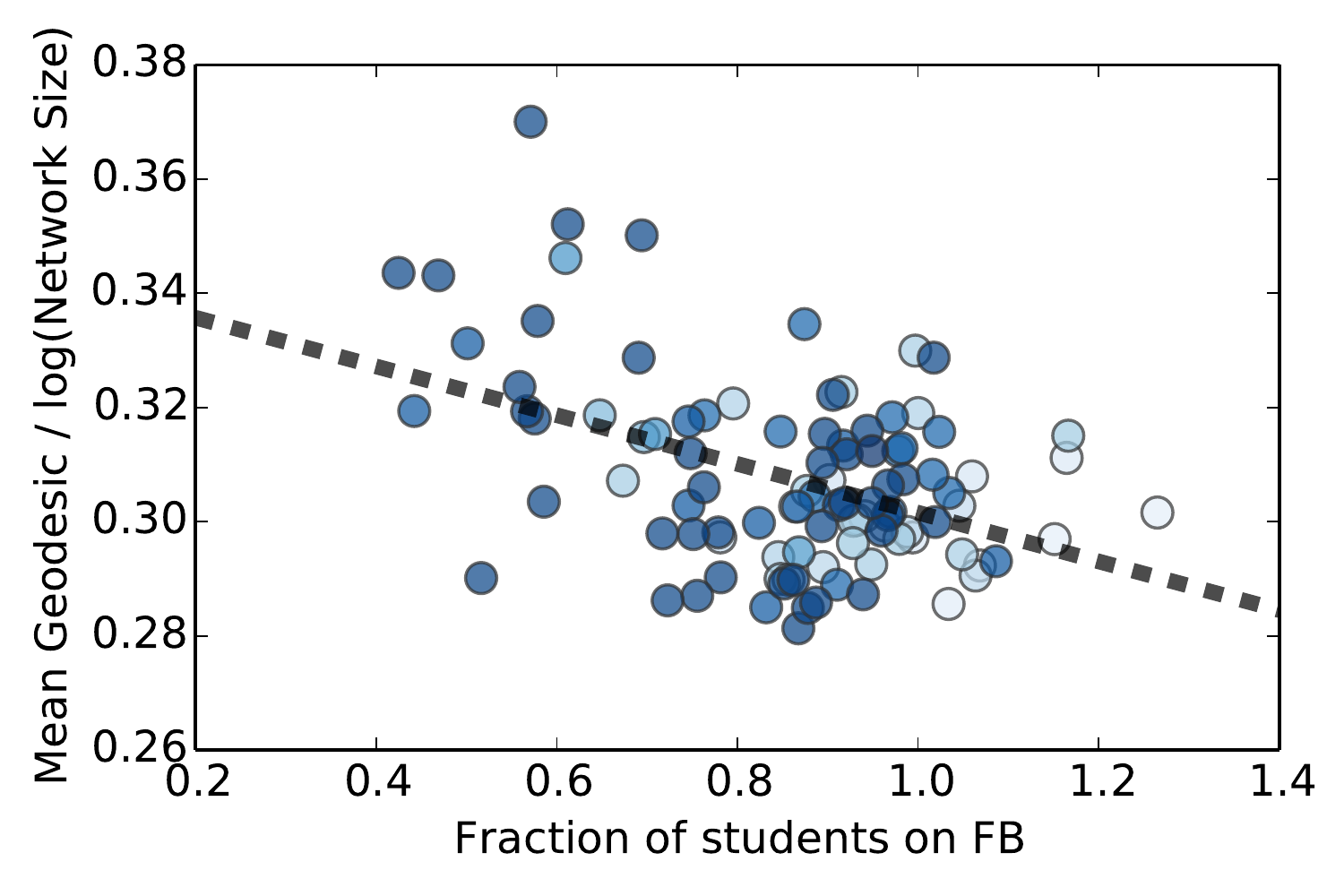}
\caption{Even after controlling for size, the mean geodesic distance decreases with adoption in undergraduate networks. Color corresponds to the vintage of the network by date added.}
\label{fig:geodesicadoption}
\end{center}
\end{figure}

As Facebook was introduced to different colleges, each school's online social network grew within a finite social space, limited by the size of the student population. The fraction of service adoption describes the relative growth in these populations and is therefore a plausible measure of the maturity of each network in this context. We expect to see more clear correlations between measures of network structure and the maturity of a networks's assembly process. (Because adoption levels are estimated only among students, we restrict these analyses to the induced subgraph among student vertices.) In Fig.~\ref{fig:sizevsage} we find a relationship between vintage and adoption. We also find that as adoption increases, the normalized mean geodesic distance, i.e., the distance divided by the overall $O(\log n)$ pattern, tends to decrease slightly (Fig.~\ref{fig:geodesicadoption}). That is, the greater the level of adoption, the shorter the paths between a pair of individuals, controlling for network size (Fig.~\ref{fig:geodesicsize}). Thus, adoption, rather than size, may be a better measure of the maturity of a network under assembly. Furthermore, this supports the two-phase developmental process, in which path lengths should grow during a sparse growth phase, and become on average shorter as the network densifies.

\begin{figure}[t!]
\begin{center}
\includegraphics[width=0.225\textwidth]{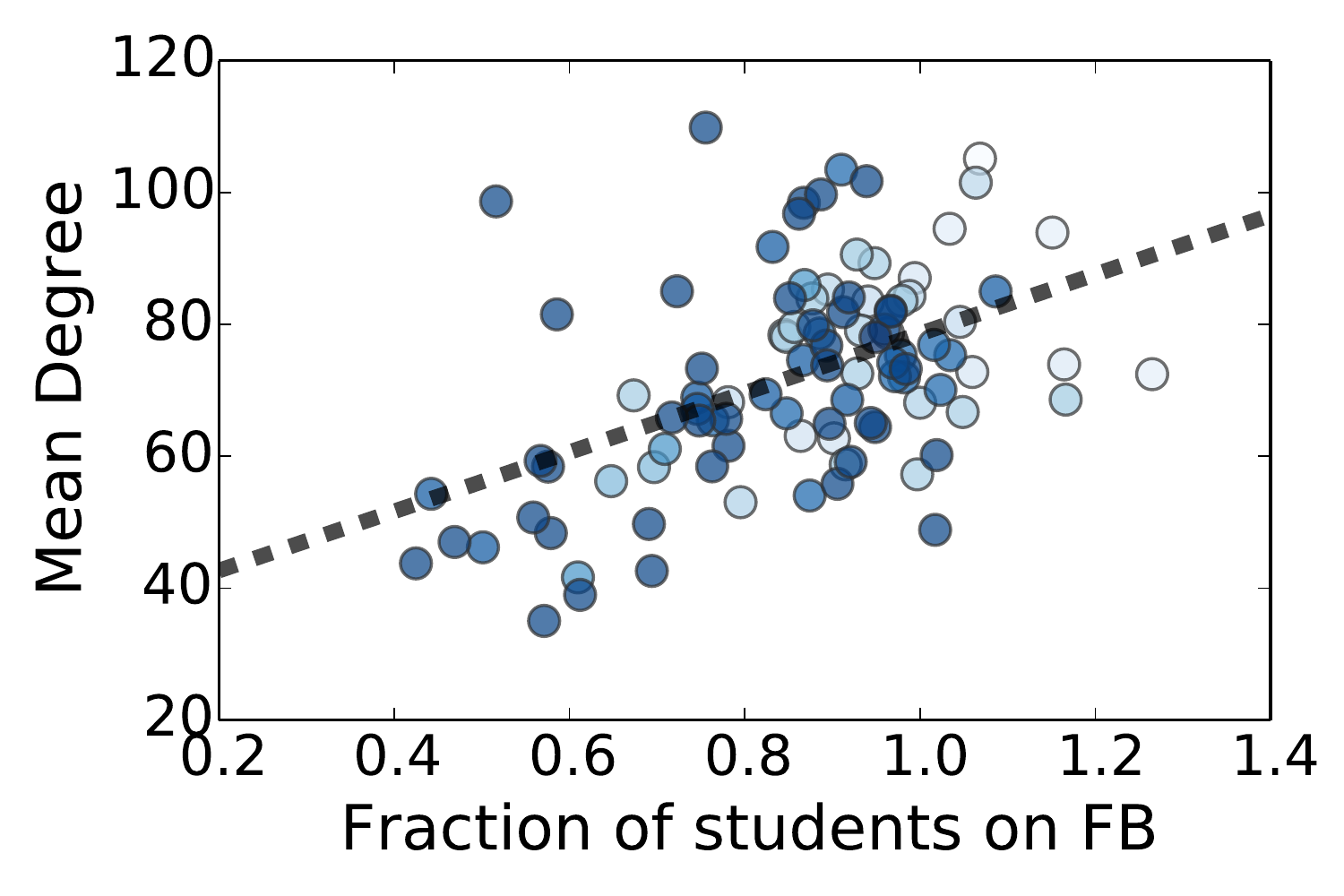}
\includegraphics[width=0.222\textwidth]{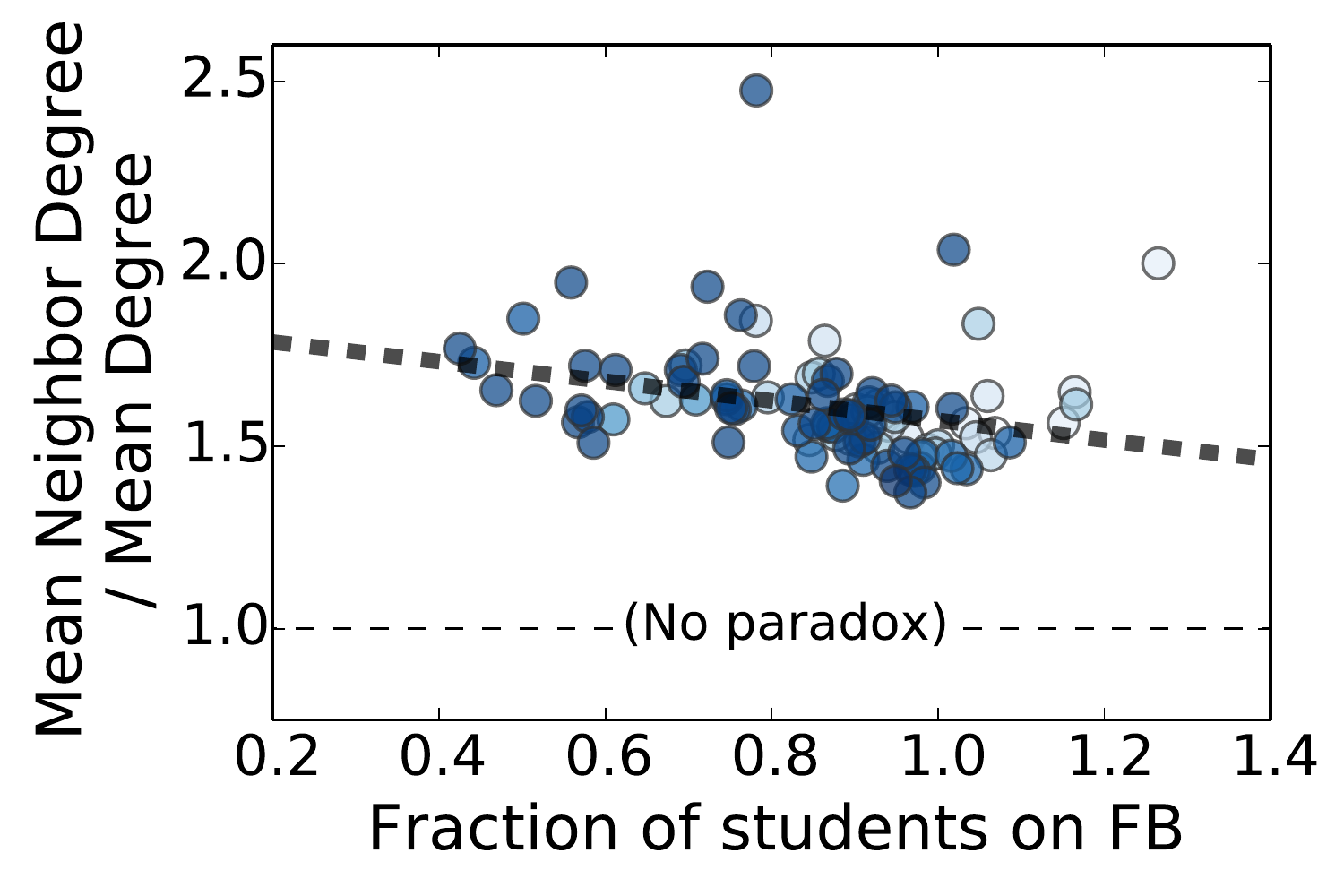}
\caption{Mean degree increases and degree distributions become less skewed in more mature networks, shown here by adoption rate. Color corresponds to the vintage of the network by date added.
}
\label{fig:friendshipparaadoption}
\end{center}
\end{figure}

The degree distribution is a network description of great interest, with social networks frequently exhibiting heavy-tailed degree distributions. A consequence of this heavy-tailedness is the unequal distribution of mean neighbor degree to mean degree~\cite{kooti2014network}. For regular graphs this ratio is one, while for all other degree distributions it is necessarily greater than one. We use this ratio as a proxy for the heavy-tailedness of the degree distribution, and find that degree distributions become less heavy-tailed as networks mature (Fig.~\ref{fig:friendshipparaadoption}). That is, even though the mean degree of a random \text{neighbor} of a vertex and the mean degree of a random vertex both tend to increase with adoption, the mean degree of a random vertex grows slightly faster as a network matures. This pattern is consistent with the two-phase developmental pattern suggested above, where an initial phase of sparse growth with many new vertices and comparatively few connections are added, and then followed by a densification phase, where new connections are mainly added between existing vertices.

Together, these results argue that network assembly is not simply network growth, or vintage, or adoption, and furthermore, that the Facebook100 networks are drawn from a single online social network assembly process. However, heterogeneity of the network assembly processes is induced by differences in network size and network adoption. The Facebook100 networks can provide useful insights into how these mechanisms interact, and heterogeneity within subpopulations of these networks can potentially reveal greater insight into the assembly mechanisms at play.

\section{Heterogeneities from\\ natural experiments}
\label{sec:HeterogeneousMaturity}

Accidents of history and the timing of our snapshot induced auspiciously observable heterogeneities in the online and offline assembly processes of our population of college networks. In this section, we examine these heterogeneities as natural experiments to explore the variability in online social network structure due to differing processes. These natural experiments are useful because they let us examine how different subpopulations of users differ in their connectivity, which lets us identify the detailed processes by which these networks assemble.

We begin by first examining basic differences among different subpopulations defined by graduating class year. We then use the timing of the arrival of freshmen on campus (in 2005, at the time of the snapshot) and the arrival of Facebook on campus (in 2004, either before and after the class of 2004 graduated) to investigate the maturity of the online social networks more precisely.  Finally, we find that the subnetworks that had less time to mature (due to environmental and historical reasons) share broad structural patterns with the university networks that had lower adoption rates. 

\begin{figure}[t!]
\begin{center}
\subfigure{\includegraphics[width=0.22\textwidth]{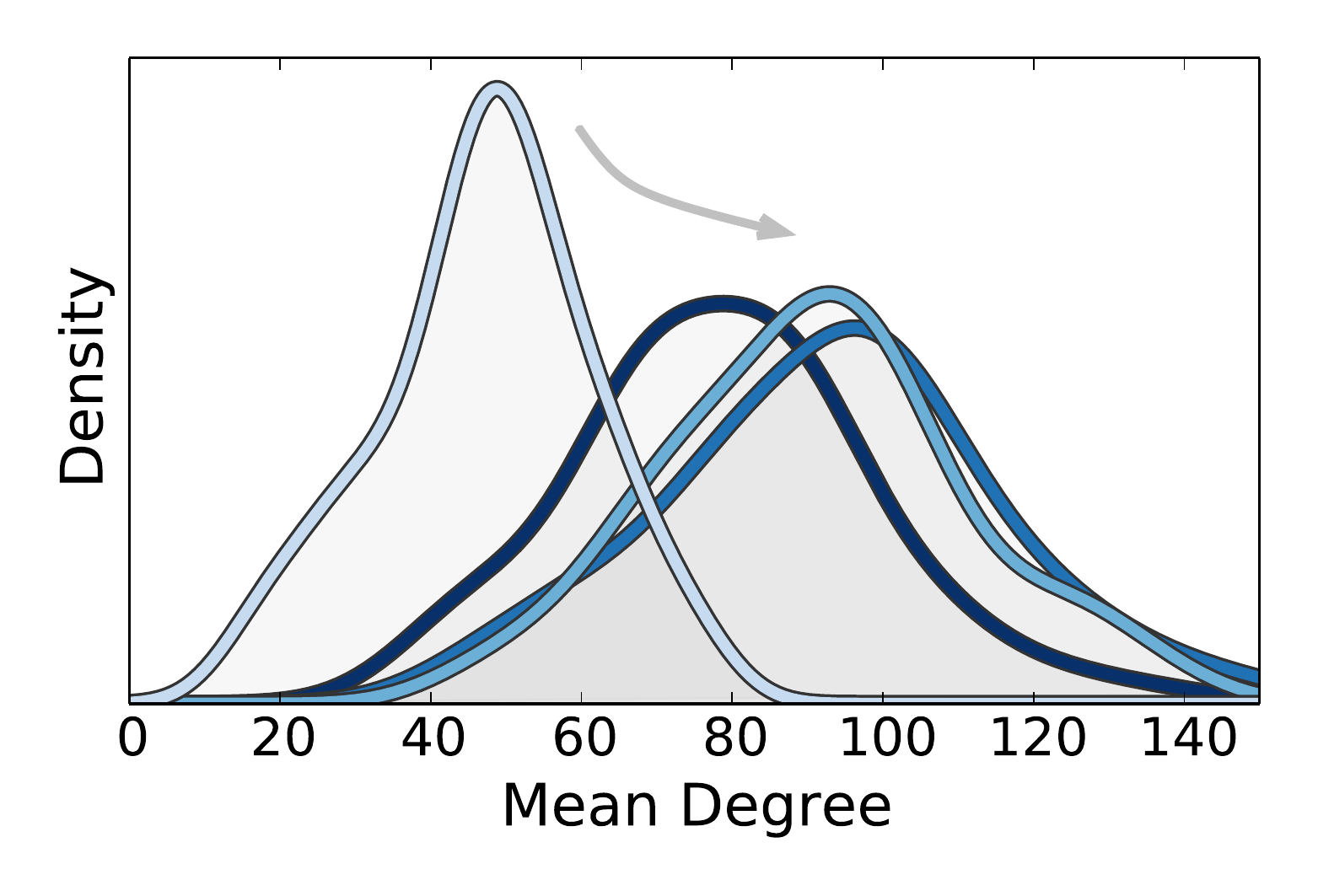}}
\subfigure{\includegraphics[width=0.22\textwidth]{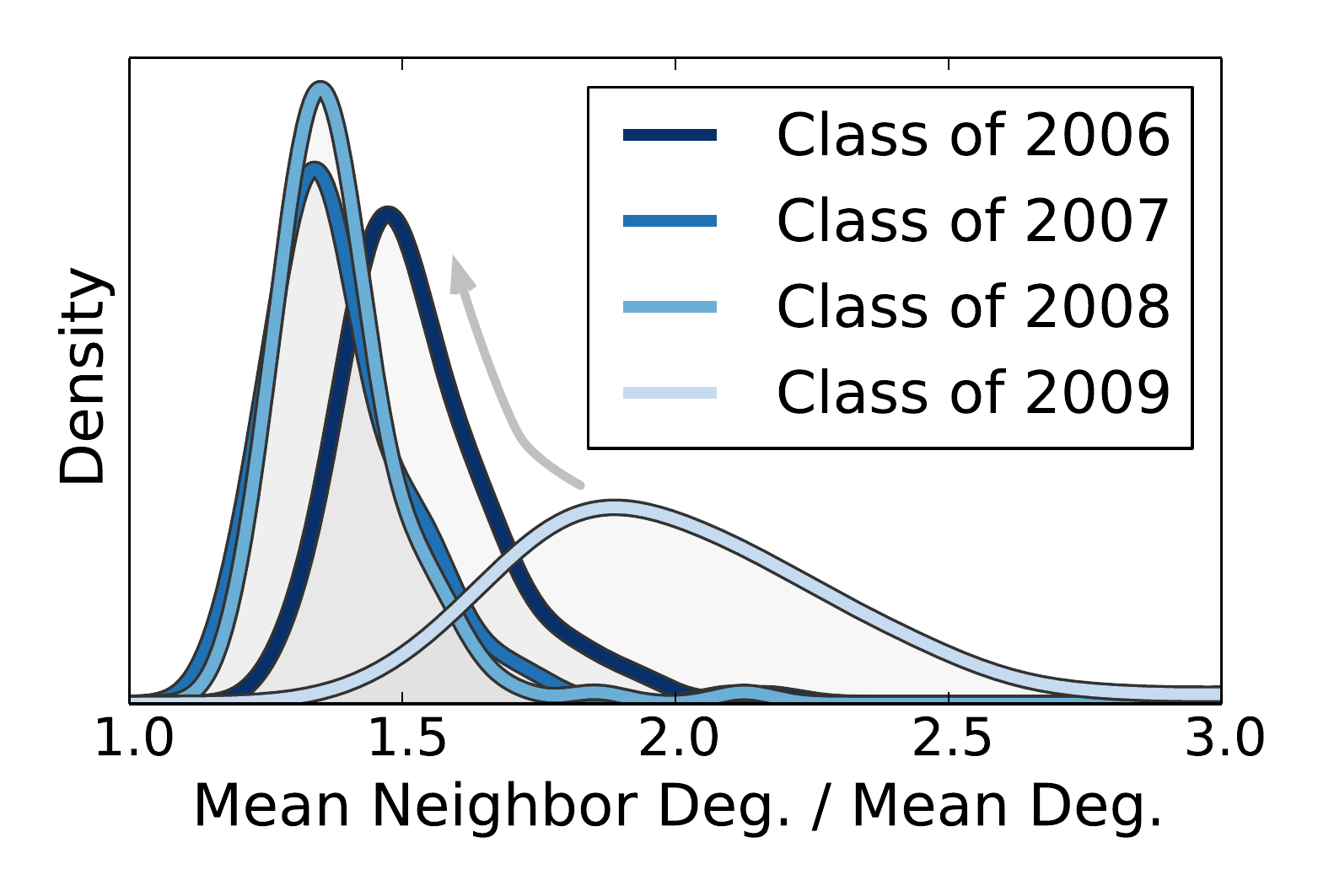}}
\subfigure{\includegraphics[width=0.22\textwidth]{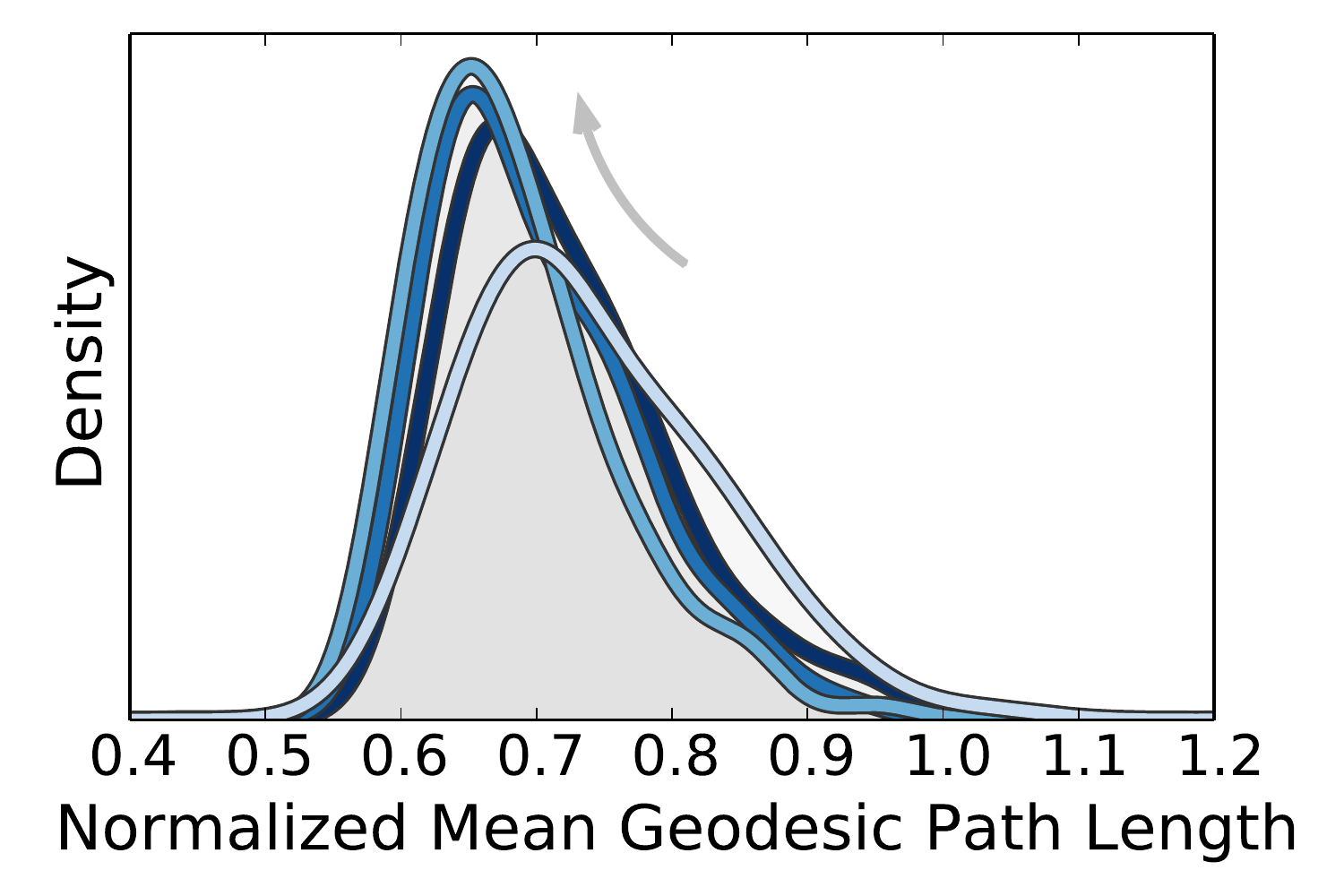}} 
\subfigure{\includegraphics[width=0.22\textwidth]{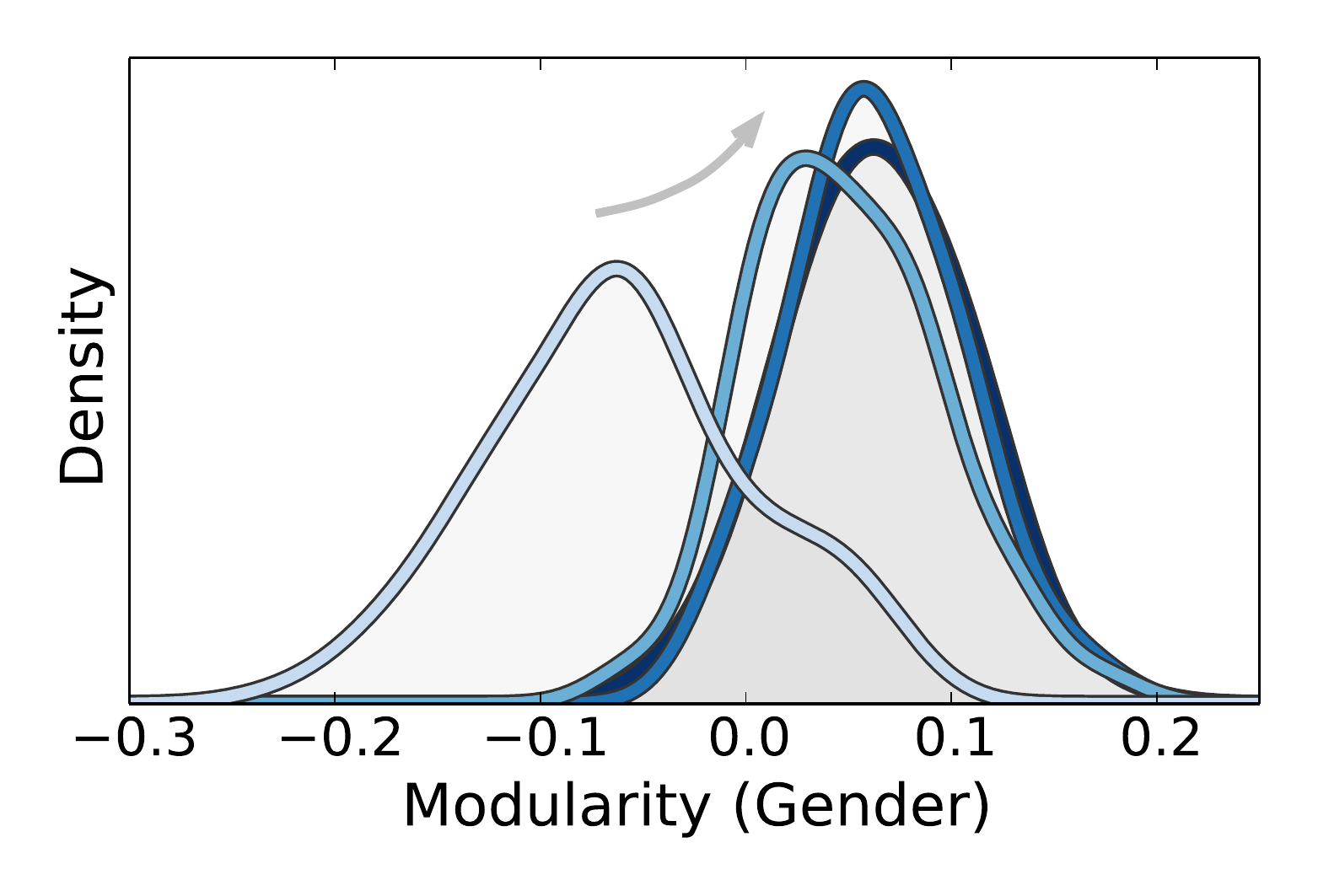}}
\subfigure{\includegraphics[width=0.22\textwidth]{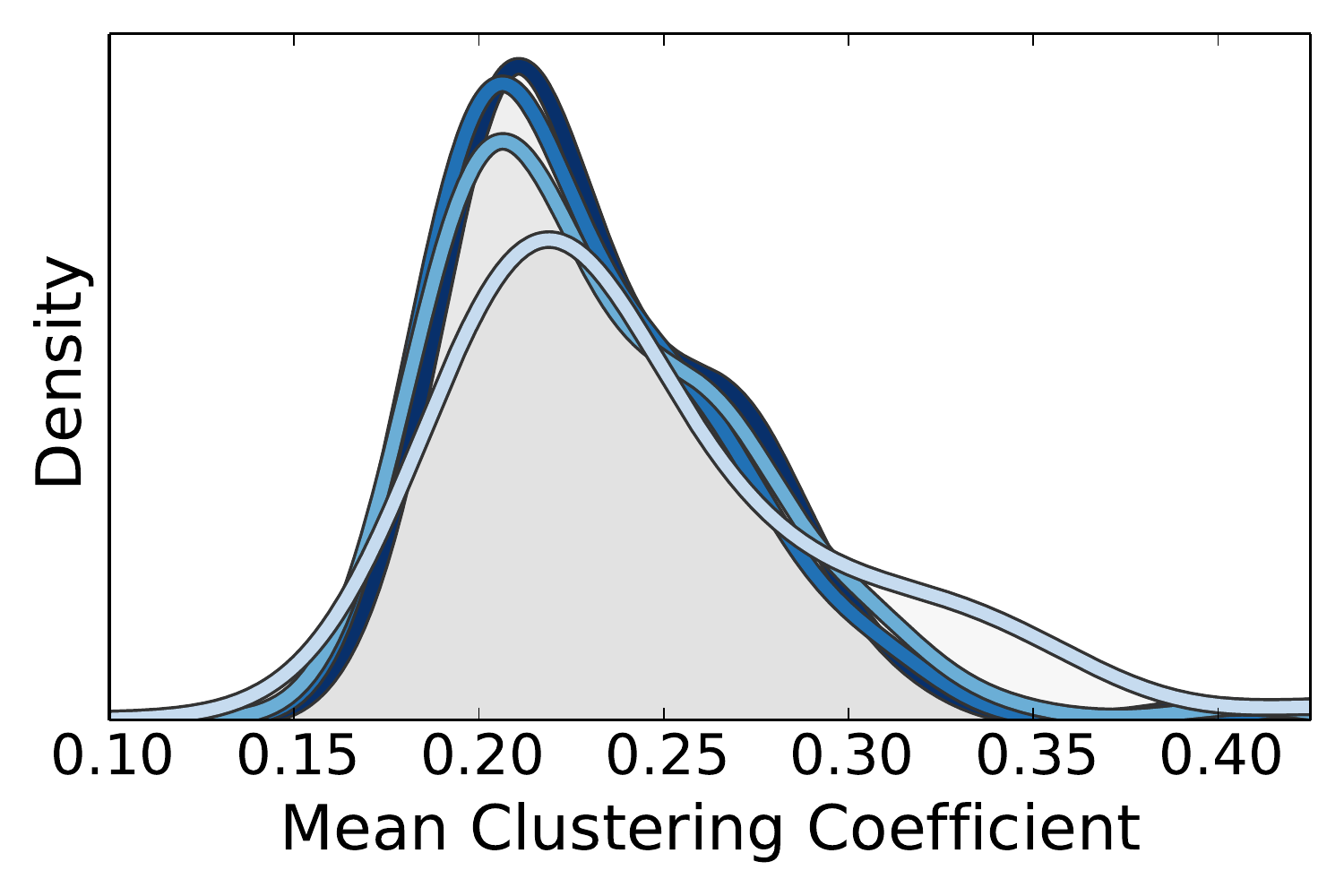}}
\subfigure{\includegraphics[width=0.22\textwidth]{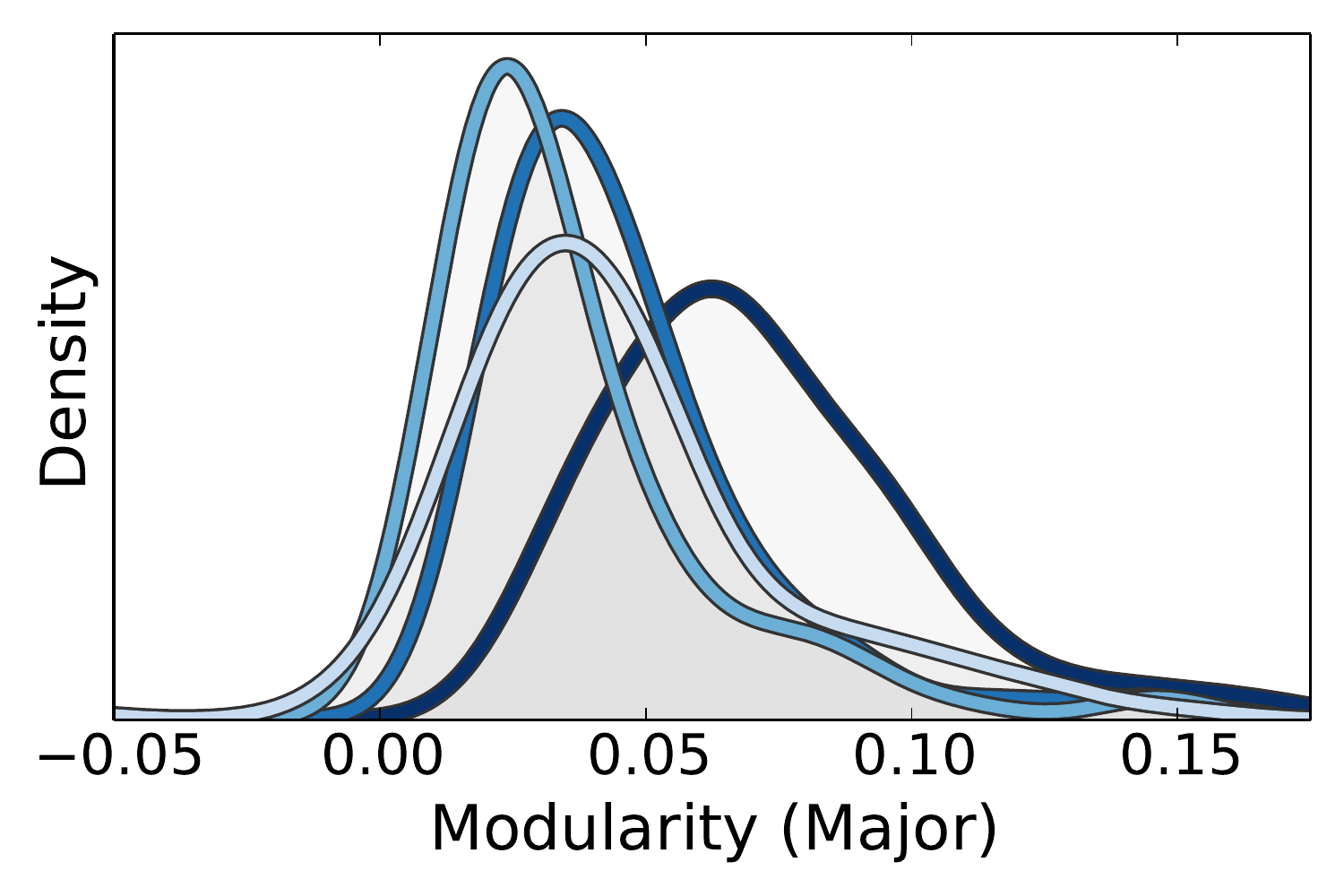}}
\caption{Distributions of undergraduate network features across the population of 100 schools, by graduating class. Distributions are visualized using kernel density estimation. Arrows move from class of 2009 to classes of 2007 and 2008, the classes with the highest adoption, when the difference between those distributions is statistically significant (two-sample KS test, $p<0.01$).
}
\label{fig:sophomorefeats}
\end{center}
\end{figure}

We first look at differences among the undergraduate population (Fig.~\ref{fig:sophomorefeats}). The classes of 2008 and 2009 arrived on campus as freshmen in the fall of 2004, at a similar time or after Facebook, and thus formed their offline and online social networks almost concurrently. Previous work found that classes with more established offline networks prior to Facebook's arrival had observable differences in behavior: survey research conducted within our sample showed that the classes of 2008 and 2009 were more likely than the classes of 2006 and 2007 to form offline friendships as a result of online friendships~\cite{ellison2007benefits}. On the other hand, for the classes of 2006, 2007, and 2008, students had access to Facebook for a similar amount of time, so these networks should have had equal opportunity to assemble. Thus, we can investigate the roles of time and offline social context among these classes.

Between the classes of 2006, 2007, and 2008, we observe that the class of 2006 has notably lower mean degree, a more skewed degree distribution, and higher modularity by major. The lower mean degree and higher skew are consistent with a less mature network, possibly due to lower engagement~\cite{tufekci2008grooming}, while the higher modularity by major suggests that these upperclassmen simply mix less across majors.

\begin{figure}[t!]  
\begin{center}
\subfigure{\includegraphics[width=0.32\textwidth]{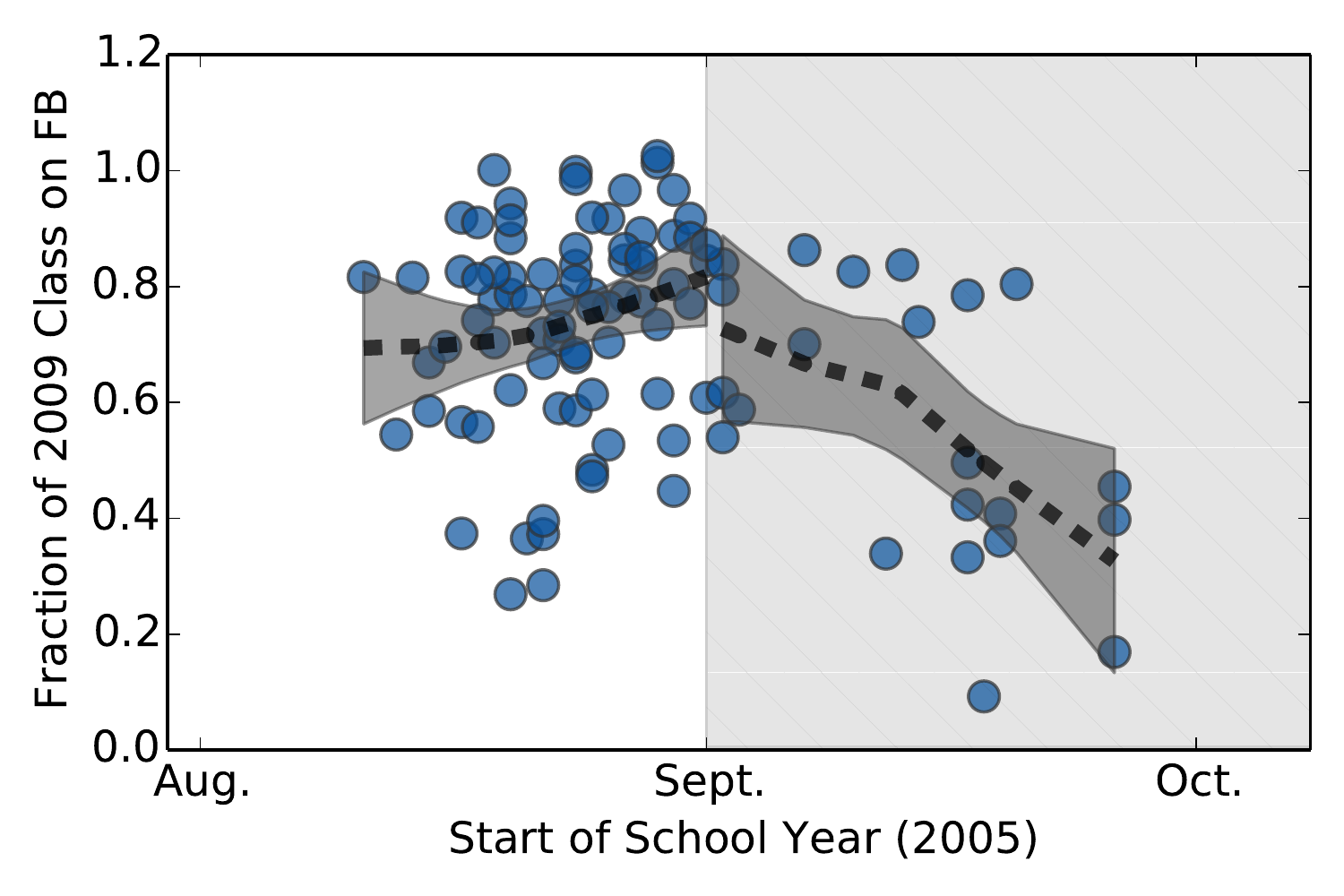}}
\subfigure{\includegraphics[width=0.22\textwidth]{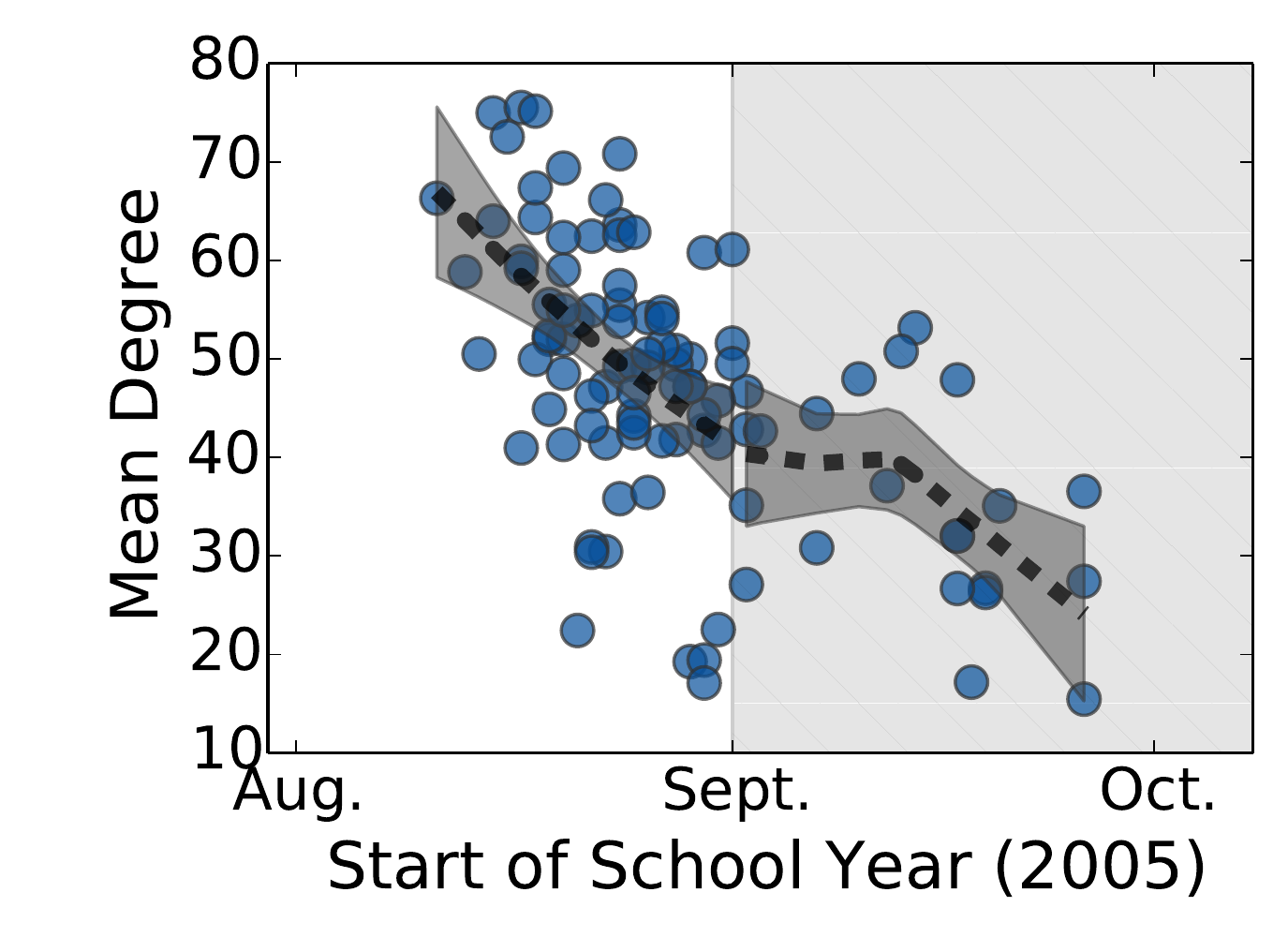}}
\subfigure{\includegraphics[width=0.22\textwidth]{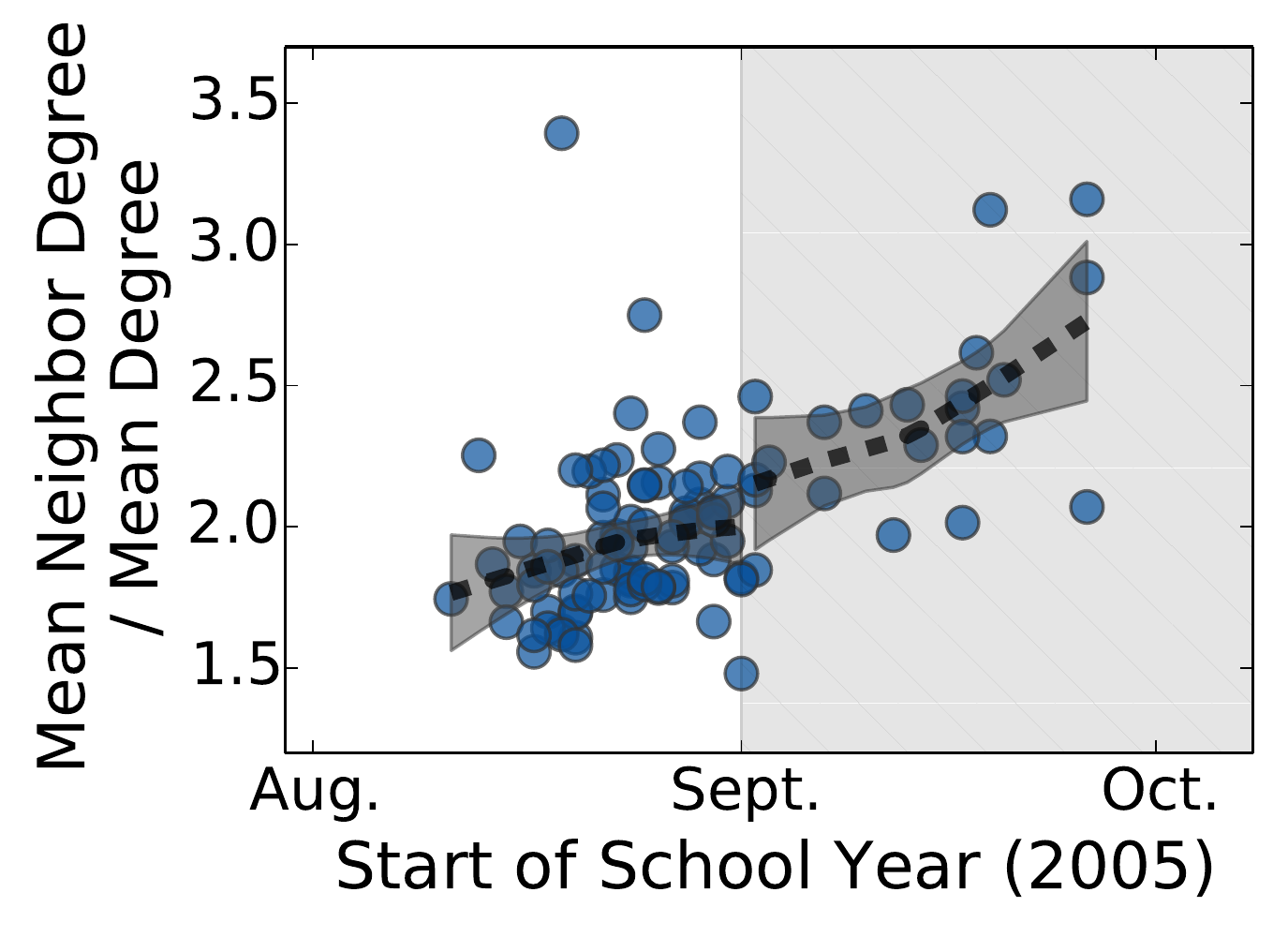}}
\subfigure{\includegraphics[width=0.22\textwidth]{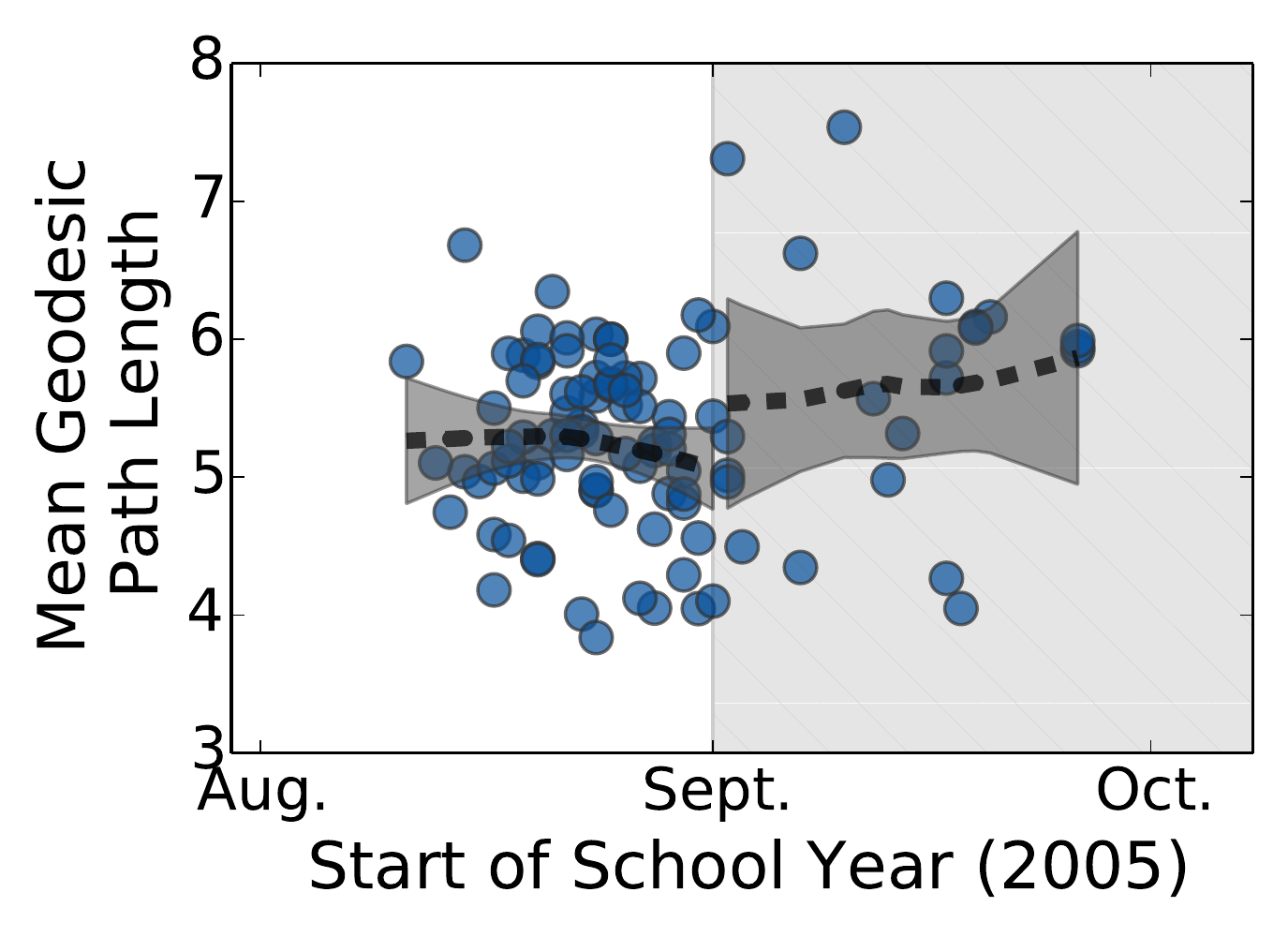}}
\subfigure{\includegraphics[width=0.22\textwidth]{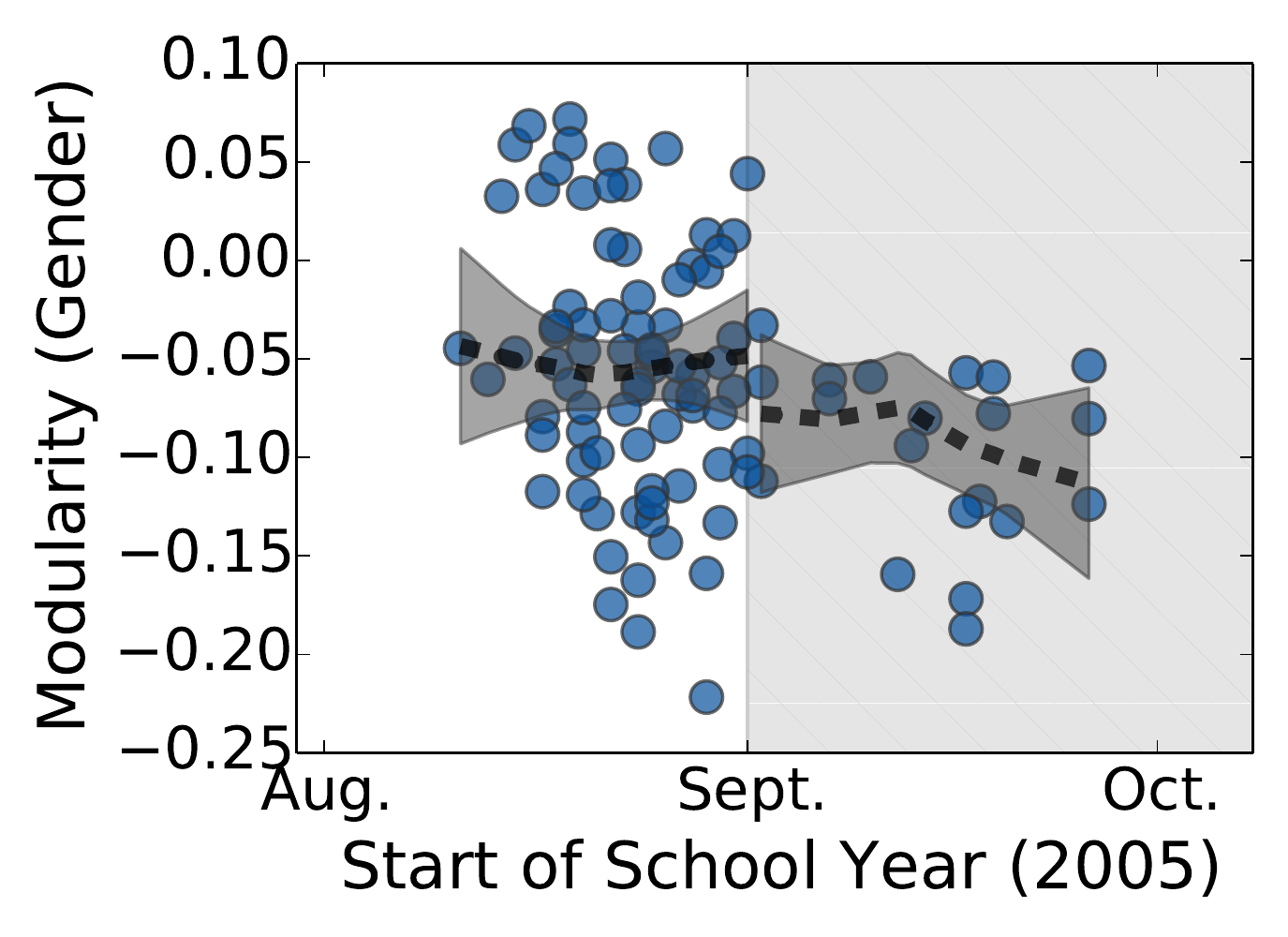}}
\caption{Network features ordered by date new students arrived on campus, August--September 2005.
The snapshot was taken in early September 2005 (gray). The dashed lines are LOESS curves over schools that began before and after September 1, shown with 95\% confidence intervals about the mean.
}
\label{fig:friendshipparafresh}
\end{center}
\end{figure}

\xhdr{Class of 2009 natural experiment} 
Across most statistics, the most strikingly different distributions are those that describe the class of 2009 networks (Fig.~\ref{fig:sophomorefeats}). The class of 2009 primarily began their undergraduate careers in the fall of 2005, when the snapshot of our data was taken. As these new students only recently gained university affiliations, the class of 2009 networks would have had the least time to develop. Notably, a fraction of these classes would have arrived on campus before the snapshot was taken, and those classes could have an offline basis for their online friendships.

Overall, the class of 2009 networks have lower average degree, more skewed degree distributions, and are disassortative by gender, whereas the older classes are assortative by gender. Studying these differences at the distributional level, it is not clear whether the differences we see in Fig.~\ref{fig:sophomorefeats} are the result of the reduced vintage of these subnetworks, with students having only joined Facebook during the summer of 2005, or some difference of assembly connected to the principally online interactions that formed these networks. Enter the natural experiment.

Students enrolling in the fall of 2005 generally obtained access to Facebook during the summer of 2005, in conjunction with obtaining university email addresses. Activity on Facebook for students not yet on campus was essentially limited to online ``social browsing'' \cite{lampe2006face}, as they possessed no offline context yet to motivate ``social searching.''  Through Internet-archaeological research, we gathered the calendar dates that incoming freshmen arrived on campus in 2005 at the 100 involved colleges to discover if and to what degree the observed differences in network structure could be connected to opportunities for offline interactions (Fig.~\ref{fig:friendshipparafresh}). We first observe a strong relationship whereby the networks for new students who have spent more time on campus---but similar amounts of time socializing online---are more mature. Students that have spent more time on campus have higher mean degree, less skewed degree distributions, as well as higher adoption overall. Interestingly, we find strong evidence for a pattern of social browsing focusing on the opposite gender: students that have spent more time physically together, and thus are more actively engaging in social search, are more gender assortative than students that have primarily interacted online.

Controlling for the size of the freshman networks, there are three data points of particular interest:\ Northeastern, Caltech, and Tulane. At Northeastern, most undergraduates are enrolled in programs that are explicitly five-year programs: that is, students identify at the outset as having a five-year graduation date. (This is in contrast to most colleges, where students enter identifying with a four-year graduation date, despite potentially longer times to completion.) For the Northeastern networks, the class of 2009 shares properties well-aligned with the second year (sophomore) students at other schools; this should be expected, as most of the members of the Northeastern class of 2009 began college in Fall 2004, not 2005. Caltech, meanwhile, is known to have an exceptional social environment among the schools in the Facebook100 dataset, as was studied closely in earlier work \cite{traud2011comparing,traud2012social}. Caltech is an outlier on almost every network metric including clustering coefficient and modularity by dorm. The structure of Tulane's class of 2005 has at play unique external events, namely the massive disruption due to Hurricane Katrina, which hit New Orleans on August 29, 2005. Tulane freshmen ultimately spent very little time physically on campus, but may have coped with this significant event by connecting through the medium of Facebook during the early days following.

\begin{figure}[t!]
\begin{center}
\subfigure{\includegraphics[width=0.22\textwidth]{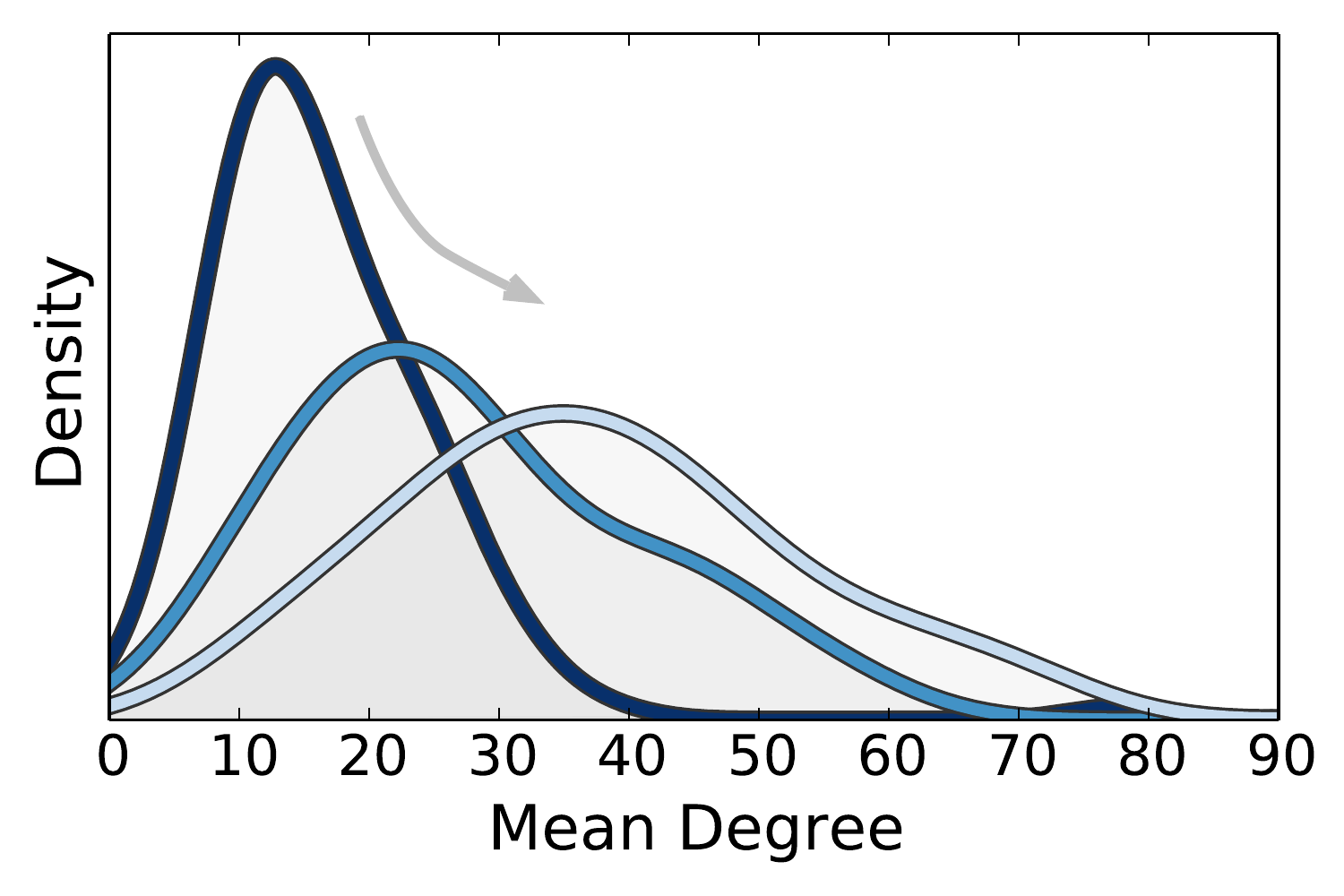} }
\subfigure{\includegraphics[width=0.22\textwidth]{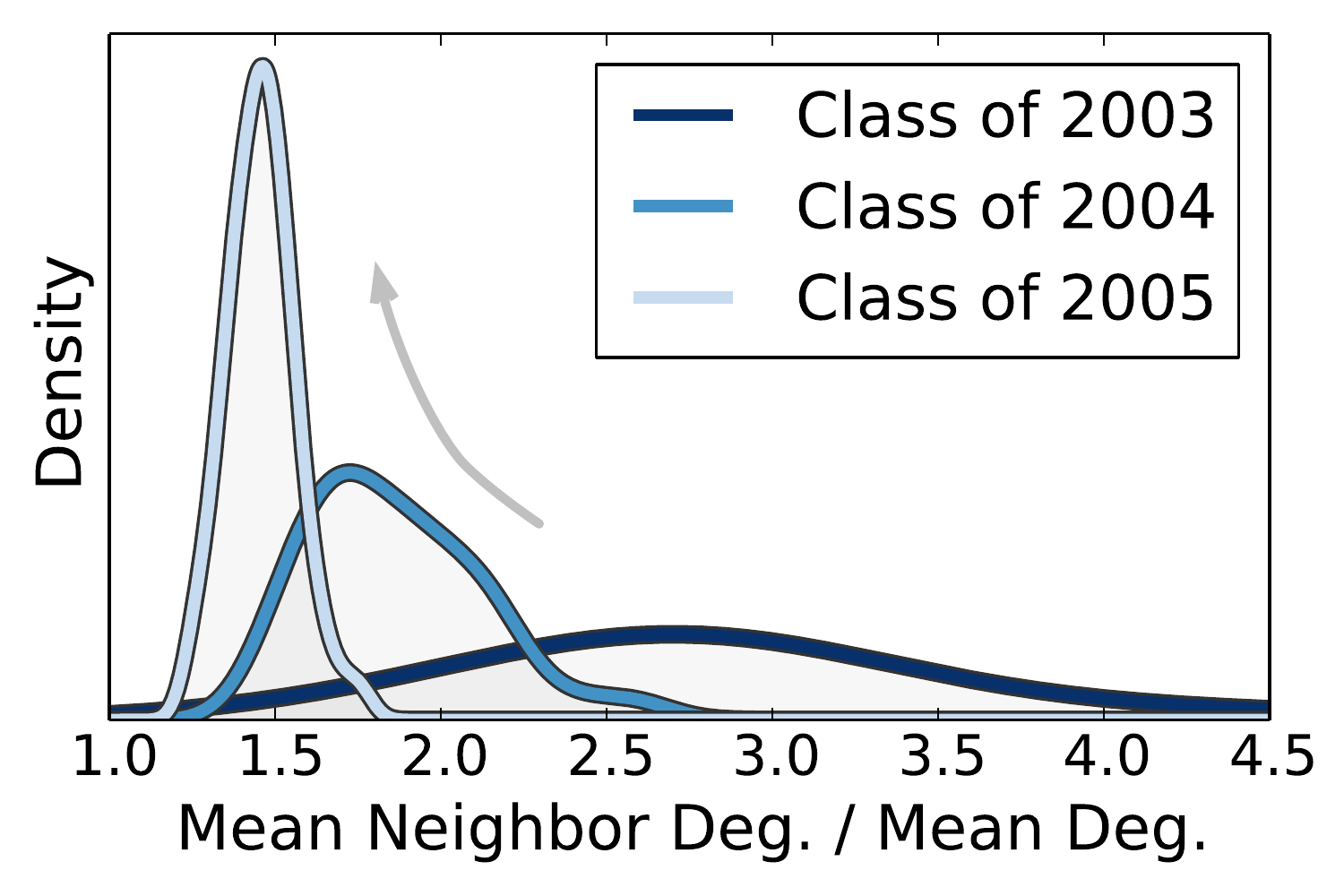}}
\subfigure{\includegraphics[width=0.22\textwidth]{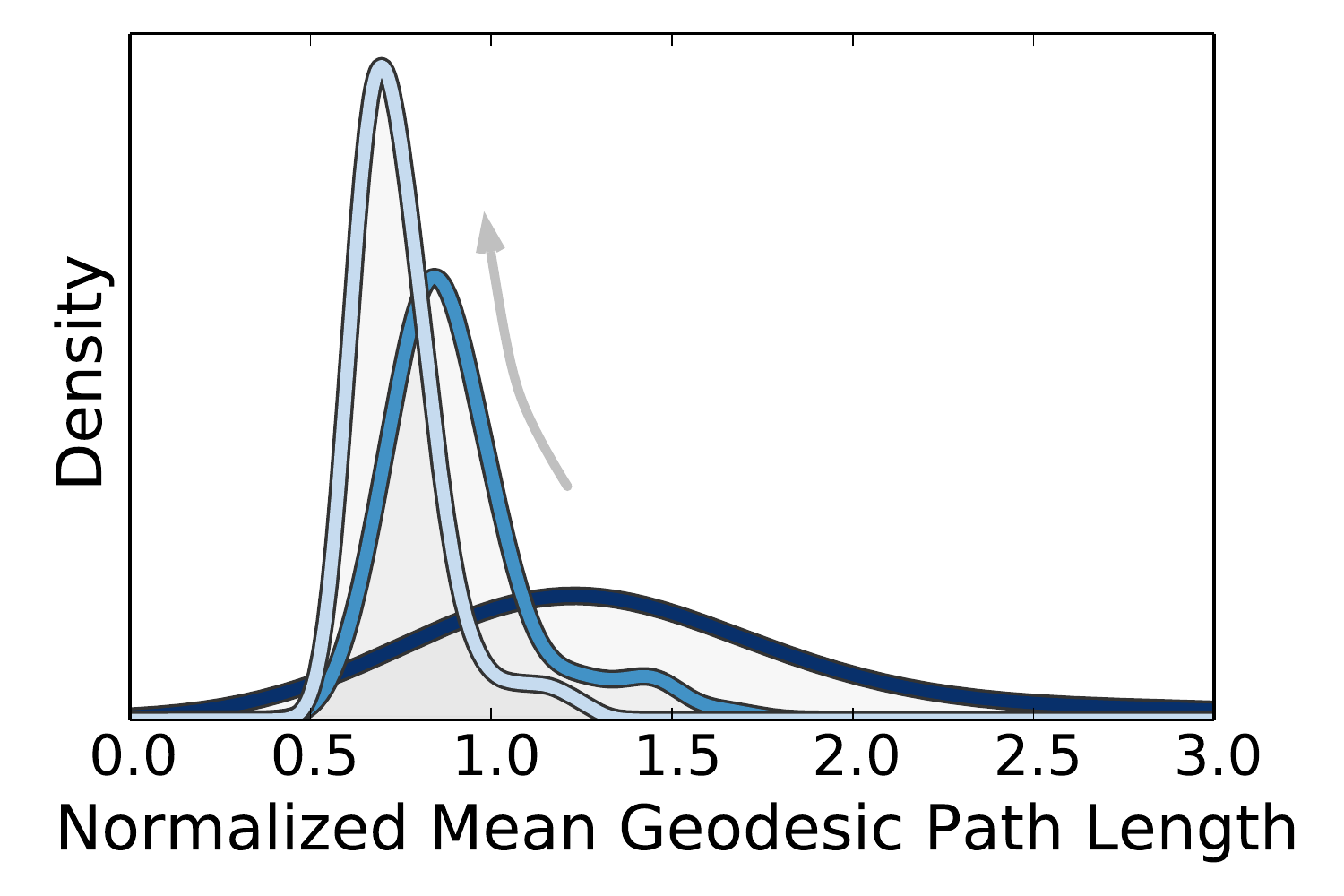}} 
\subfigure{\includegraphics[width=0.22\textwidth]{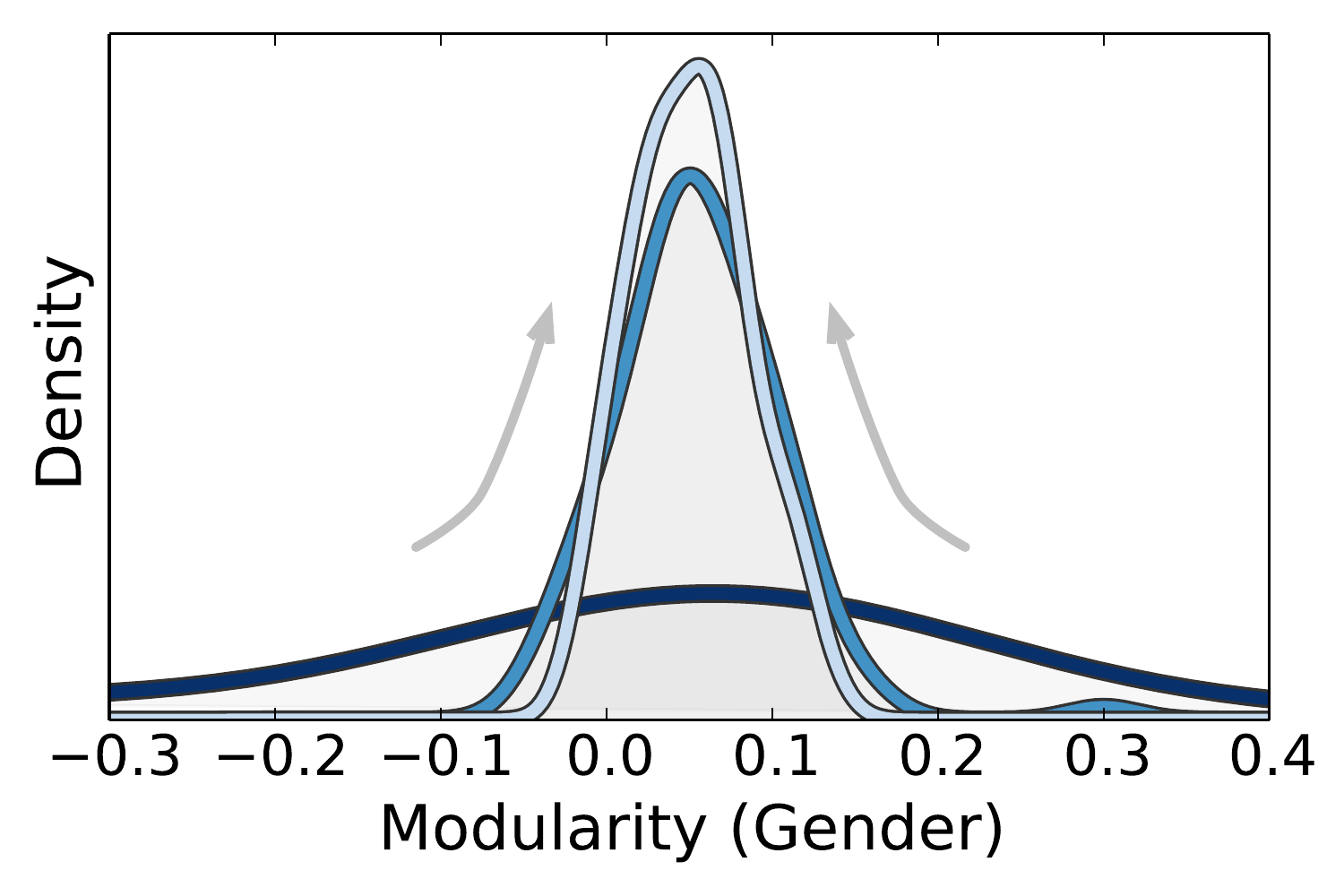}}
\subfigure{\includegraphics[width=0.22\textwidth]{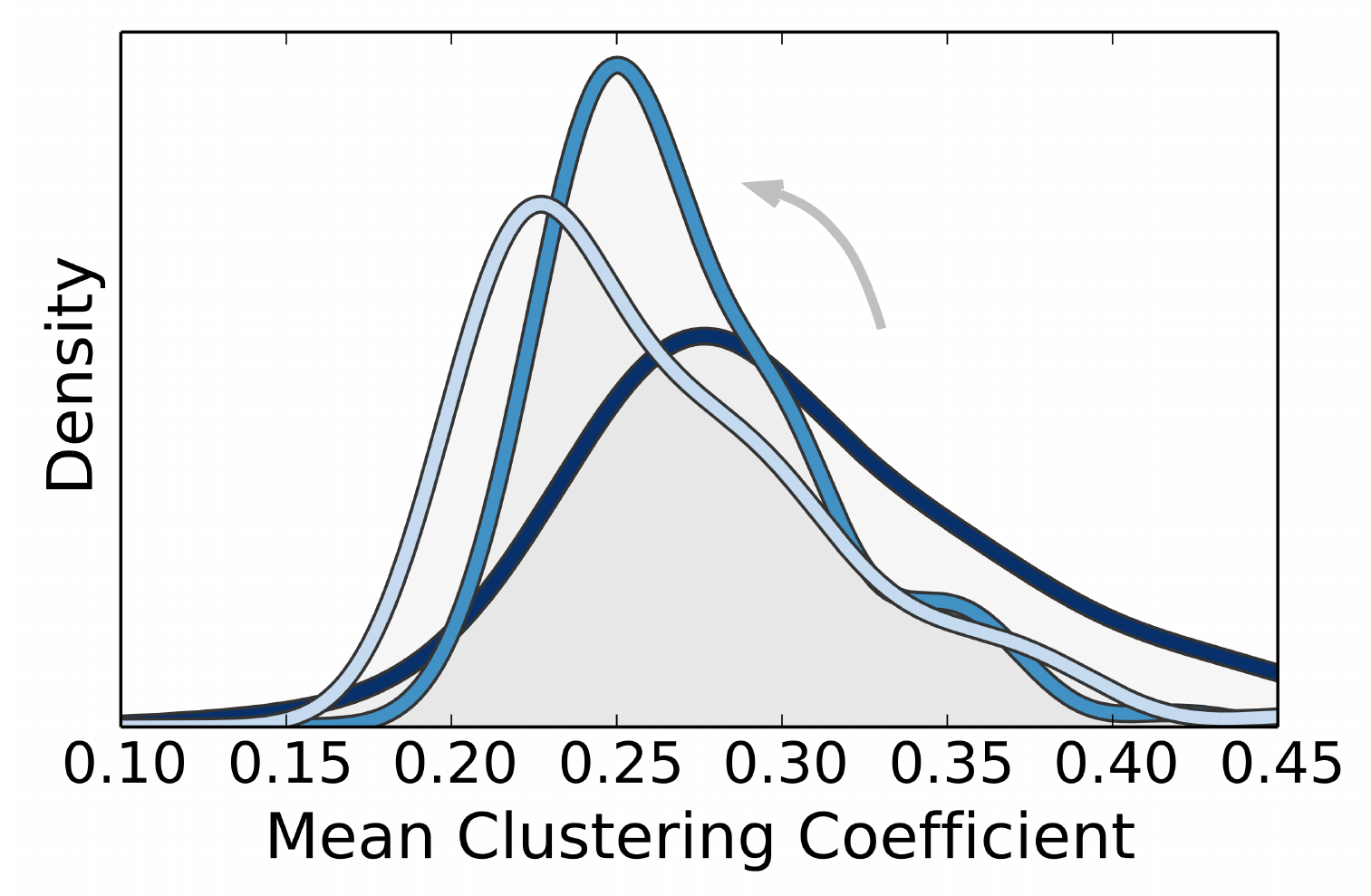}}
\subfigure{\includegraphics[width=0.22\textwidth]{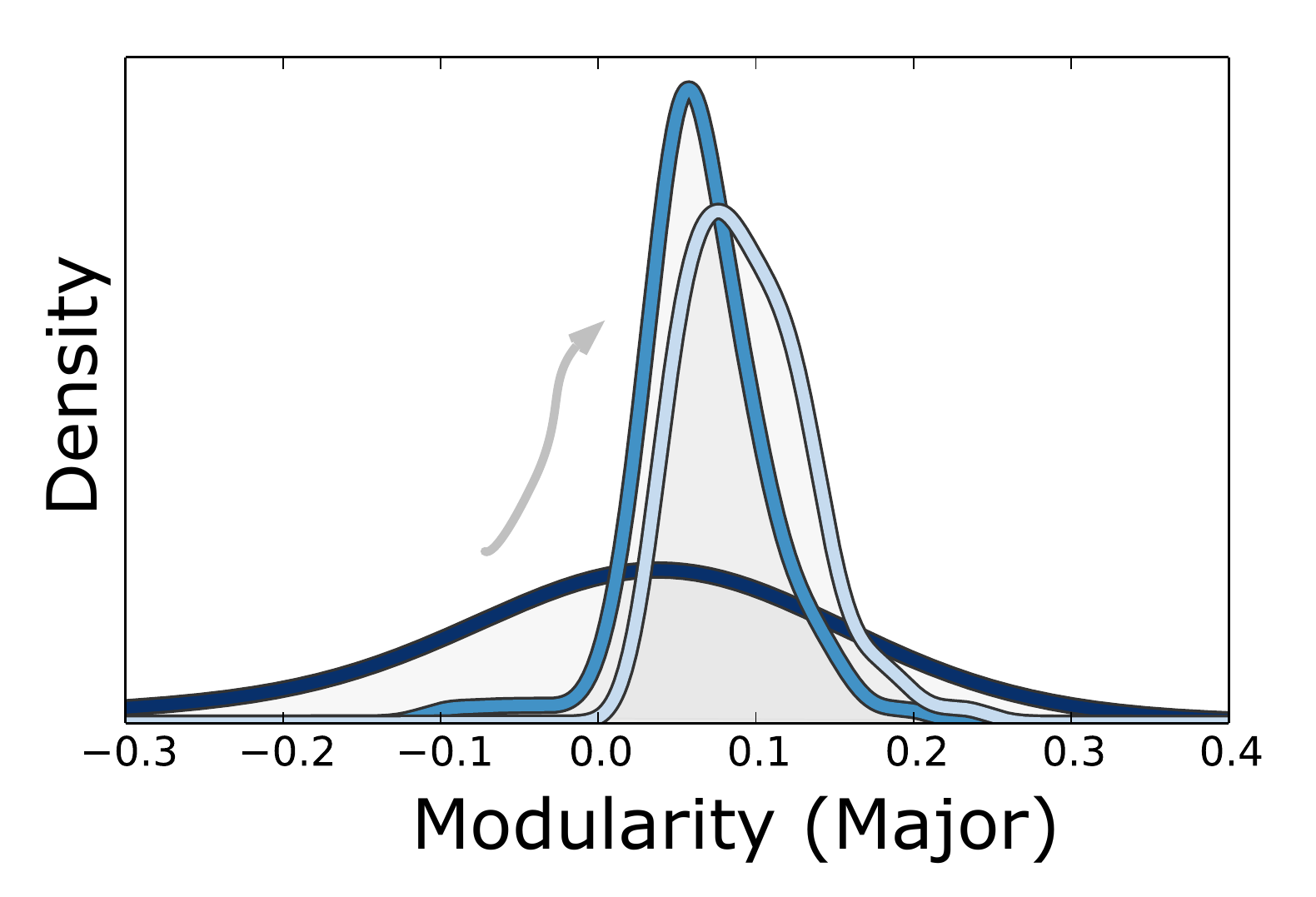}}
\caption{
Distributions of alumni network features across the population of 100 schools, by graduating class. Distributions are visualized using kernel density estimation. Arrows move from the class of 2003 (lowest adoption) to the class of 2005 (highest), when the difference between those distributions is statistically significant (two-sample KS test, $p<0.01$).
}
\label{fig:alumvsUGfeatures}
\end{center}
\end{figure}

\begin{figure}[t!]
\begin{center}
\subfigure{\includegraphics[width=0.43\textwidth]{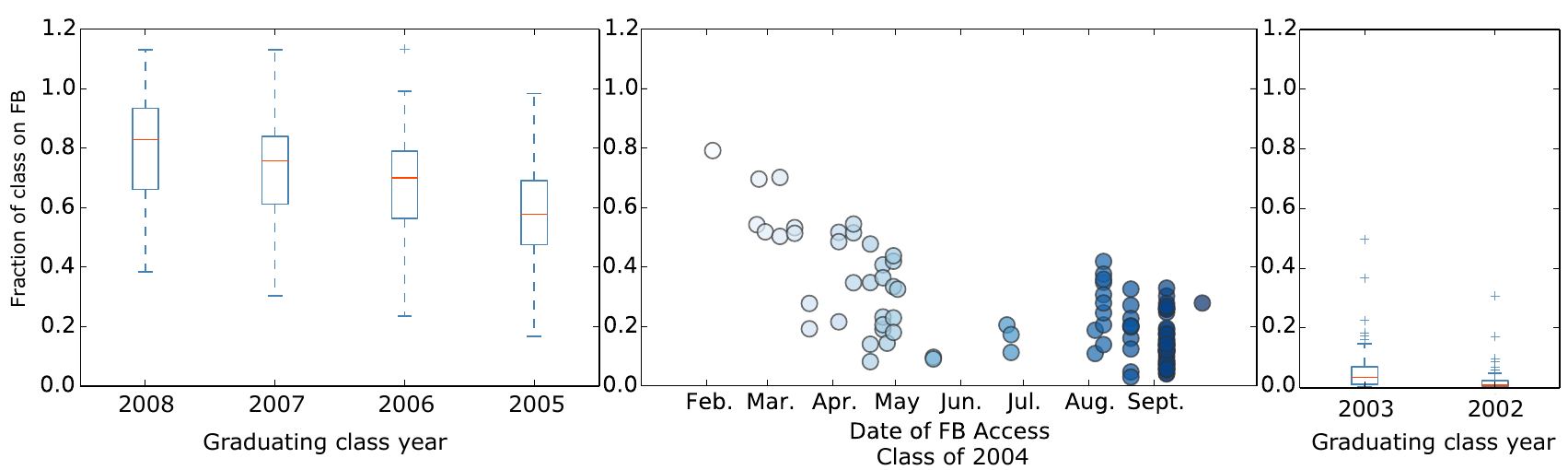}} 
\subfigure{\includegraphics[width=0.225\textwidth]{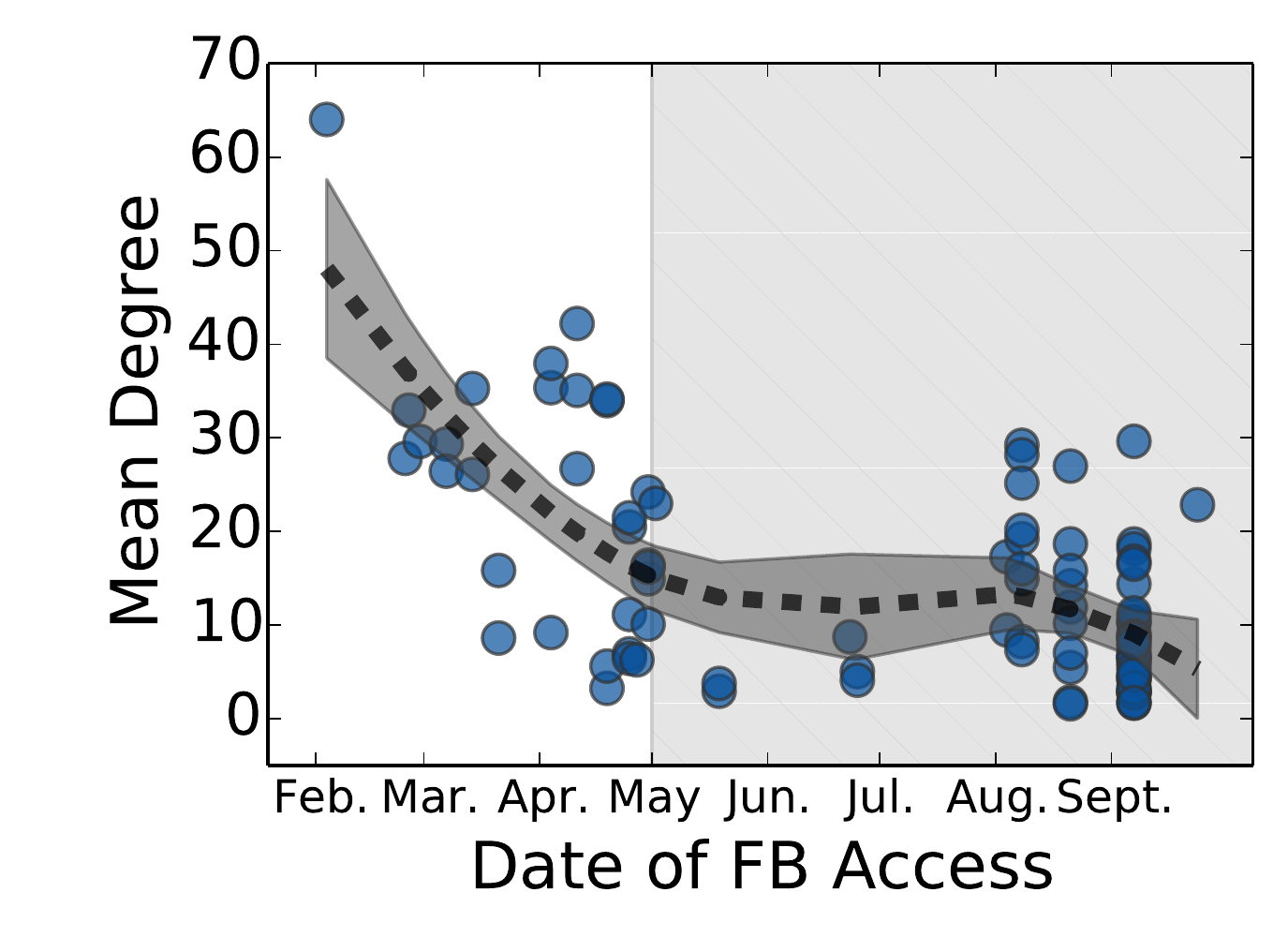}}
\subfigure{\includegraphics[width=0.225\textwidth]{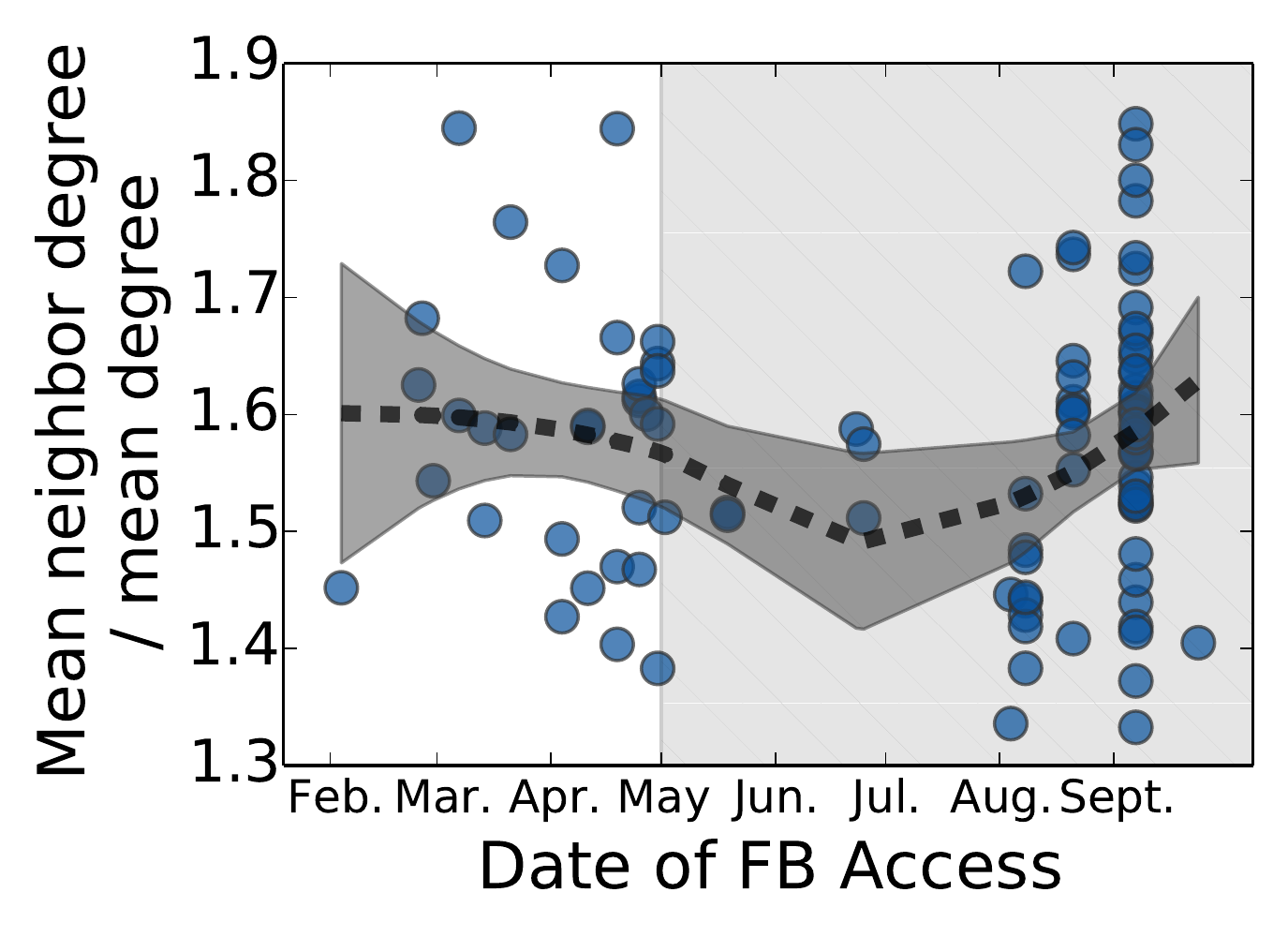}}
\subfigure{\includegraphics[width=0.225\textwidth]{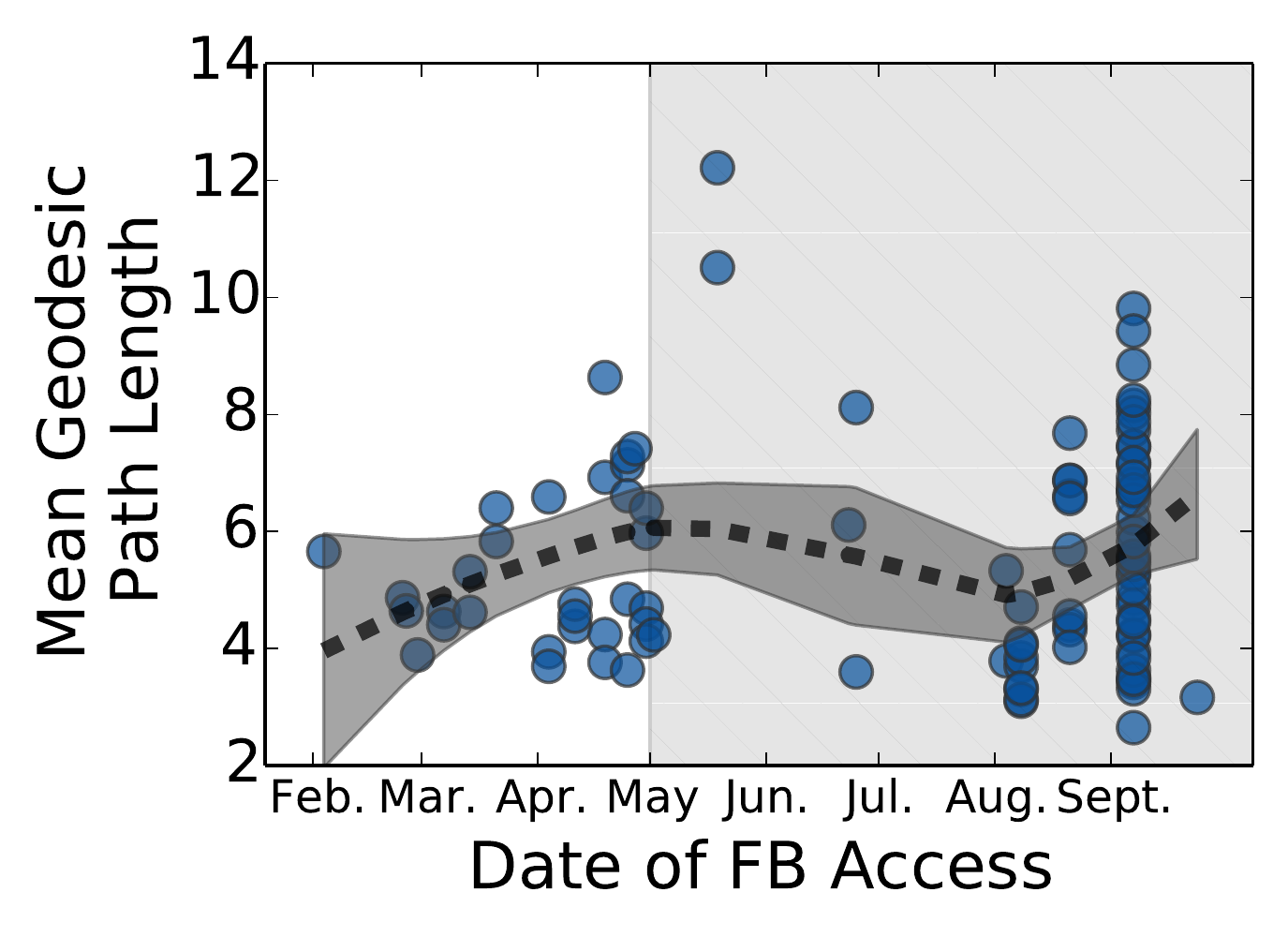}}
\subfigure{\includegraphics[width=0.225\textwidth]{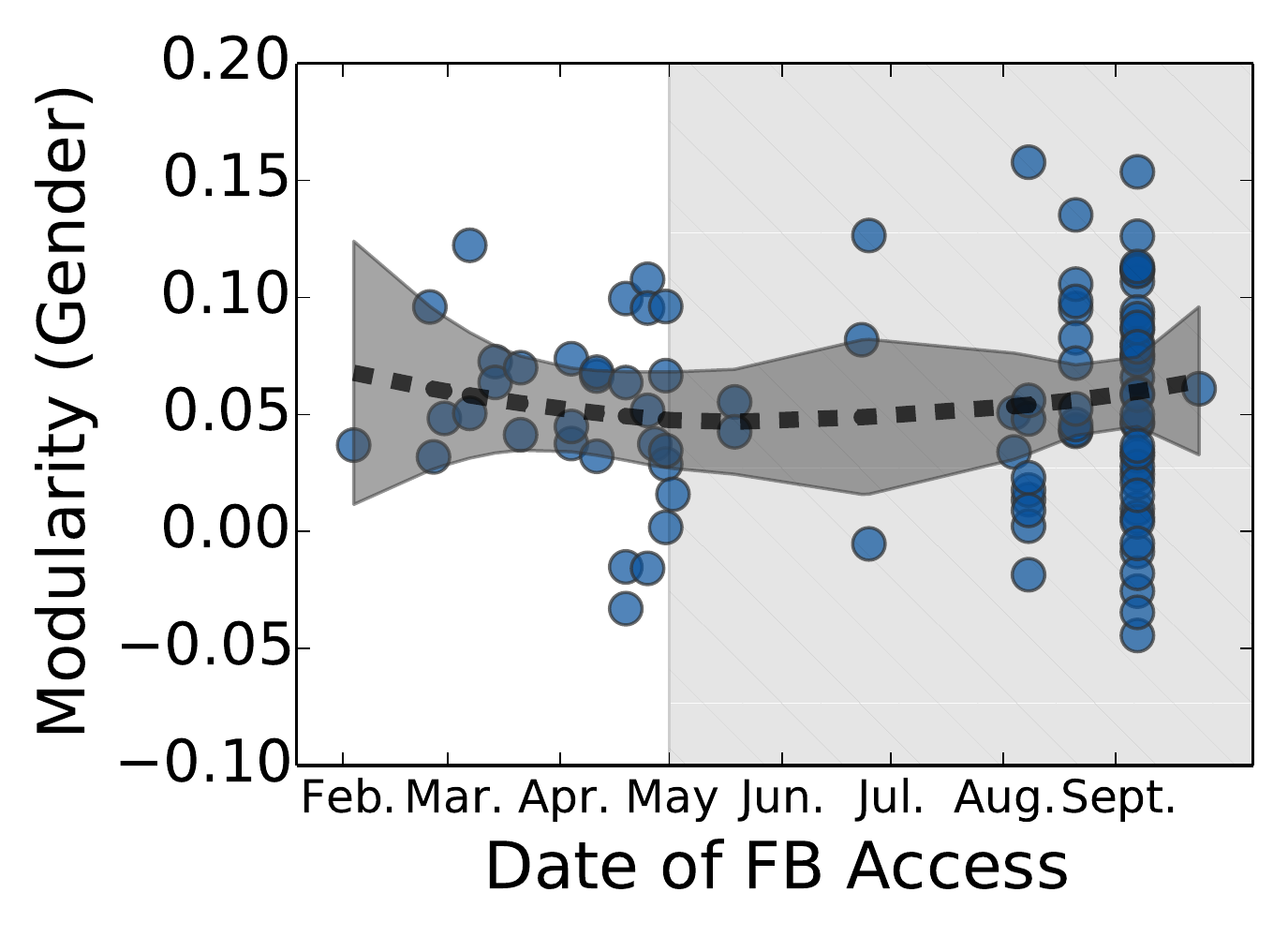}}
\caption{(top) Network adoption for class years. The boxes are bound by the 25th and 75th percentiles, and the center line is the median.  (top center) Network adoption for each university network by the class of 2004, ordered and shaded by date the university gained access to Facebook. (below)
Network properties for the class of 2004 by date of access to Facebook. The shaded region separates classes that graduated prior to gaining access to Facebook, and the dashed lines are LOESS curves, shown with 95\% confidence intervals about the mean.
} 
\label{fig:classof2004properties}
\end{center}
\end{figure}

\xhdr{Class of 2004 natural experiment} 
Shifting our focus to the opposite temporal end of the dataset, the alumni in our sample reflect a diversity of social, spatial and cultural settings, and notably lacked the opportunity for closed mixing within university campuses. In Fig.~\ref{fig:alumvsUGfeatures}, we consider three graduating classes of students: 2003, 2004, and 2005, which in sum comprise on average 84.4\% of the alumni users at the time of the snapshot. (Less than 5\% of alumni have observable earlier class years; 11.4\% of the alumni do not report their class year.) We first investigate differences between these three classes, of which the class of 2005 spent almost a full year with Facebook while colocated on campus; some of the class of 2004 gained access to Facebook before graduating (Fig.~\ref{fig:FBtimeline}), a distinction we will explore more deeply next; and the class of 2003 only having gained access to Facebook after graduation. We analyze the induced subgraphs of these alumni classes, and find that the more recent alumni networks are more mature, and furthermore that the class of 2004 network appears to represent a maturity level intermediate to the class of 2003 and 2005. This smooth transition suggests that the university environment induces additional online assembly of the offline social networks being captured.

The graduating class of 2004 primarily finished their undergraduate careers during May and June of 2004. Concurrently, Facebook was spreading to increasingly many campuses, with students at Harvard (id$=$1) graduating after several months on Facebook, and the University of California San Diego (id$=$34) after just a few weeks. Of the 100 colleges in the sample, 66 did not gain access to Facebook until after the class of 2004 graduated, so those new alumni would no longer share the university environment when they joined.

Again using Internet archaeology---primarily via the Internet Archive, the Spring 2004 Media Kit from TheFacebook LLC, and student newspapers---we collected the dates that universities joined Facebook in order to tease apart the effects of the university environment on the early growth of the Facebook network. Across the different school networks, the class of 2004 student populations have approximately constant demographics, and the first 34 schools are comparable by size, public/private status, and geography compared to the remaining 66 (Figs.~\ref{fig:FBtimeline},~\ref{fig:sizevsage}). Thus, other things being equal, we can examine the impact of the arrival of Facebook on the network assembly of the class of 2004.

At the time of the snapshot, over a year past most students' graduation and granted access to Facebook, adoption still tracks strongly with the arrival time of Facebook (Fig.~\ref{fig:classof2004properties}). We also find that mean degree correlates with arrival time, both of which suggest that the offline and cohesive social environment played a role in the rate at which these networks grew. Other variables did not exhibit a strong trend throughout this transition. This negative result suggests that the class of 2004 networks were of relatively constant maturity level. Arguably, this maturity level interpolates between the classes of 2003 (whose network assembly was almost exclusively outside of the college environment) and 2005 (whose graduating students were able to connect while on campus) (Fig.~\ref{fig:alumvsUGfeatures}), whereas the size of the network was largely determined by the amount of time in a shared offline context. This suggests that the initial transition into alumni status realized a similar level of complexity of existing offline social structures, as opposed to the sharp transition exhibited among freshmen arriving on campus, with a discrete start time and novel social connections. This suggests that the type of shared offline context plays a significant role in the trajectory of network assemblies.

\section{Discussion and Conclusions\label{sec:discussion}}

The large size, early rise, and storied history of Facebook make it a model system for studying the processes and patterns of online social network assembly, i.e., the complicated and heterogeneous changes these networks undergo as they mature. The Facebook100 networks capture a special part of this history---the first 20 months, the first 100 colleges, and the first one million users---which allows us to investigate the early stages of assembly. Our analysis sheds new light on the extent to which a network's assembly is driven by simple growth, how a network's structure changes as it matures, how network structure varies with adoption, and how the connectivity patterns of different groups of users tends to converge, at different rates, on similar end states. 

Each of these results depended on our using a population of social graphs to measure distributions of structural statistics, which allowed us to better estimate the natural variability of network structure produced by the underlying social processes. In contrast, many other studies rely on a single network instance, which makes it difficult to identify whether some pattern reflects a general insight or a special case. Many questions and tasks in the analysis of networks would benefit from this kind of population approach.

Applied to the Facebook100 data, this approach revealed several novel insights into the assembly of online social networks. First, these graphs exhibit a clear $O(\log n)$ dependence for the mean geodesic distance (Fig.~\ref{fig:geodesicsize}). This pattern agrees closely with conventional wisdom, which is largely drawn from classic results in random graph theory, but it defies recent claims about general ``densification laws,'' which predict shrinking rather than growing distances. These results are not, in fact, contradictory, and instead suggest that online assembly proceeds through two distinct phases.

Initially, a network grows via sparsification, adding many new vertices from the extant population and a relatively smaller number of connections among them. For early Facebook, each time a new college joined, or a new class arrived on campus, this phase started anew within that population and proceeded as the adoption rate rose from zero. The second phase begins once the network has expanded to include a large fraction of the available population. Then, assembly transitions into a densification pattern, adding many connections among existing vertices and a smaller number of completely new vertices. Of note, these two phases can be seen as corresponding to the growth and saturation phases of logistic growth within a finite population~\cite{barrat2008dynamical}.

Past work on distances in the large-scale Facebook social network~\cite{backstrom2012four} corroborates our finding:\ the mean geodesic distance between users peaked in 2008 and subsequently shrank, illustrating a transition into a densification pattern around that time. Between its opening to the general population in 2006 and 2008, Facebook was expanding rapidly into new populations, and our findings imply that its large-scale structure grew according to a sparsification pattern. The 2008 transition to densification implies that Facebook's expansion into new populations began to slow then, allowing continued link formation to begin to densify the network.

We find further evidence for this same two-phase pattern within the Facebook100 networks, distributed across different subpopulations, which experience network assembly at different rates but toward similar end states. By combining these networks with additional information about Facebook adoption rates, and college graduation and matriculation dates, we leveraged two natural experiments within these networks to show how structure varied between students on and off campus, between students of different graduating years, and between alumni and current students. Each of these analyses showed a consistent behavior:\ the longer a subpopulation had access to Facebook, especially for students on campus, the greater its level of adoption. As adoption increases we see distances shrink, degrees increase, and degree distributions becomes less heavy tailed. 

This model would predict that just before Facebook opened up to the general population in 2006, the network structure within each of its college subnetworks was very mature, having reached high levels of adoption. Opening up to a wider range of users, however, moved the system as a whole back into the sparsification phase. As Facebook spread into this large and unadopted population, its diameter expanded and its degree distribution became more heavy-tailed, before transitioning back into the densification phase, as a greatly enlarged system, in 2008.

The specific processes by which online social networks assemble are also implicated by our results, which sheds new light on several understudied questions about networks. The online assembly process described above tends to sample offline individuals and relations~\cite{schoenebeck2013potential}, a pattern supported by social surveys of users at the time~\cite{tufekci2008grooming}. Online social networks that specifically reflect such offline relationships are thus different than those based on mainly online interactions. For instance, consider assortativity by gender among new students (Fig.~\ref{fig:friendshipparafresh}):\ those who had not yet arrived on campus tended to connect with students of the opposite gender. In contrast, those on campus tended to connect with those of the same gender, which is the pattern observed among older students already on campus. That is, the former group did not have the offline social interactions to ground their behavior in reality, and thus treated Facebook very differently---apparently, like a dating website---than on campus students embroiled in the rich offline social milieu of college life.

Looking forward, it seems clear that designing or modifying online social networks is a task best done with a detailed understanding of how different social factors and processes influence the particular trajectory that assembly takes, both at the level of individual users and at the level of the entire network. That is, human behavior is not independent of the design of these systems, and designs are likely to be more effective and more useful if they are informed by an understanding of their impact on the long-term structure and function of these networks. The study of online social network assembly promises to shed new light on these tradeoffs.

\section{Acknowledgments}
This work was supported by the 
NSF Graduate Research Fellowship award no.\ DGE 1144083 (AZJ) and the Butcher Foundation (SFW). The authors thank Mason A.\ Porter and Eric Kelsic for providing the Facebook100 data and Leto Peel for useful discussions.

\newpage
\begin{appendix}

For reference, we include a subset of our data below. The university names, network index, and dates each university gained access to Facebook are in Table~\ref{appxtab1}. The estimated arrival dates of the class of 2009 to campus are shown in Table ~\ref{appxtab2}.
For additional sources and complete methodology, we refer the reader to \texttt{http://azjacobs.com/fb100/}.

\begin{table*}[t]
\centering
\begin{tabular}{|rll|rll|}
\hline
FB100 Index & Name & Date Joined & FB100 Index & Name & Date Joined \\
\hline
1 & Harvard & 2/4/2004 & 51 & South Florida & 8/21/2004 \\
2 & Columbia & 2/25/2004 & 52 & Central Florida & 8/21/2004 \\
3 & Stanford & 2/26/2004 & 53 & Florida State & 8/21/2004 \\
4 & Yale & 2/29/2004 & 54 & GWU & 8/21/2004 \\
5 & Cornell & 3/7/2004 & 55 & Johns Hopkins & 8/21/2004 \\
6 & Dartmouth & 3/7/2004 & 56 & Syracuse & 8/22/2004 \\
7 & UPenn & 3/14/2004 & 57 & Notre Dame & 8/22/2004 \\
8 & MIT & 3/14/2004 & 58 & Maryland & 8/22/2004 \\
9 & NYU & 3/21/2004 & 59 & Maine & 9/7/2004 \\
10 & BU & 3/21/2004 & 60 & Smith & 9/7/2004 \\
\hline
11 & Brown & 4/4/2004 & 61 & UC Irvine & 9/7/2004 \\
12 & Princeton & 4/4/2004 & 62 & Villanova & 9/7/2004 \\
13 & UC Berkeley & 4/4/2004 & 63 & Virginia Tech & 9/7/2004 \\
14 & Duke & 4/11/2004 & 64 & UC Riverside & 9/7/2004 \\
15 & Georgetown & 4/11/2004 & 65 & Cal Poly & 9/7/2004 \\
16 & UVA & 4/11/2004 & 66 & Mississippi & 9/7/2004 \\
17 & BC & 4/19/2004 & 67 & Michigan Tech & 9/7/2004 \\
18 & Tufts & 4/19/2004 & 68 & UCSC & 9/7/2004 \\
19 & Northeastern & 4/19/2004 & 69 & Indiana & 9/7/2004 \\
20 & Illinois & 4/19/2004 & 70 & Vermont & 9/7/2004 \\
\hline
21 & Florida & 4/25/2004 & 71 & Auburn & 9/7/2004 \\
22 & Wellesley & 4/25/2004 & 72 & U San Fran & 9/7/2004 \\
23 & Michigan & 4/25/2004 & 73 & Wake Forest & 9/7/2004 \\
24 & Michigan State & 4/25/2004 & 74 & Santa Clara & 9/7/2004 \\
25 & Northwestern & 4/25/2004 & 75 & American & 9/7/2004 \\
26 & UCLA & 4/27/2004 & 76 & Haverford & 9/7/2004 \\
27 & Emory & 4/30/2004 & 77 & William \& Mary & 9/7/2004 \\
28 & UNC & 4/30/2004 & 78 & Miami & 9/7/2004 \\
29 & Tulane & 4/30/2004 & 79 & James Madison & 9/7/2004 \\
30 & UChicago & 4/30/2004 & 80 & UT Austin & 9/7/2004 \\
\hline
31 & Rice & 4/30/2004 & 81 & Simmons & 9/7/2004 \\
32 & WashU & 5/2/2004 & 82 & Binghamton & 9/7/2004 \\
33 & UC Davis & 5/20/2004 & 83 & Temple & 9/7/2004 \\
34 & UC San Diego & 5/20/2004 & 84 & Texas A\&M & 9/7/2004 \\
35 & USC & 6/23/2004 & 85 & Vassar & 9/7/2004 \\
36 & Caltech & 6/25/2004 & 86 & Pepperdine & 9/7/2004 \\
37 & UC Santa Barbara & 6/25/2004 & 87 & Wisconsin & 9/7/2004 \\
38 & Rochester & 8/4/2004 & 88 & Colgate & 9/7/2004 \\
39 & Bucknell & 8/4/2004 & 89 & Rutgers & 9/7/2004 \\
40 & Williams & 8/8/2004 & 90 & Howard & 9/7/2004 \\
\hline
41 & Amherst & 8/8/2004 & 91 & UConn & 9/7/2004 \\
42 & Swarthmore & 8/8/2004 & 92 & UMass & 9/7/2004 \\
43 & Wesleyan & 8/8/2004 & 93 & Baylor & 9/7/2004 \\
44 & Oberlin & 8/8/2004 & 94 & Penn State & 9/7/2004 \\
45 & Middlebury & 8/8/2004 & 95 & Tennessee & 9/7/2004 \\
46 & Hamilton & 8/8/2004 & 96 & Lehigh & 9/7/2004 \\
47 & Bowdoin & 8/8/2004 & 97 & Oklahoma & 9/7/2004 \\
48 & Vanderbilt & 8/21/2004 & 98 & Reed & 9/7/2004 \\
49 & Carnegie Mellon & 8/21/2004 & 99 & Brandeis & 9/7/2004 \\
50 & Georgia & 8/21/2004 & 100 & Trinity & 9/24/2004 \\
\hline
\end{tabular}
\caption{The calendar date that thefacebook become available to students at each of the first 100 colleges. 
The principle sources for this data are (i) Thefacebook LLC's ``Spring 2004 Media Kit'', which lists the dates
for the first 20 colleges, and (ii) snapshots of the landing page for thefacebook.com as recorded by the Internet Archive (archive.org).
Exact dates were discernible for 30 schools. When exact dates were not discernible, upper bounds (the latest possible date) were used. 
Our sources identified 84 of the 100 schools to within a window of at most 3 days. 
The two schools (Rochester and Bucknell) with the least certain dates are known to fall within a
window of 9 days, so may be up to 9 days earlier than listed here. For additional sources and complete methodology, see \texttt{http://azjacobs.com/fb100/}.}
\label{appxtab1}
\end{table*}

\begin{table*}[h]
\centering
\begin{tabular}{|rll|rll|}
\hline
FB100 Index & Name & 2005 Orientation & FB100 Index & Name & 2005 Orientation \\
\hline
1 & Harvard & 9/10/2005 & 51 & South Florida & 8/22/2005 \\
2 & Columbia & 8/29/2005 & 52 & Central Florida & 8/17/2005 \\
3 & Stanford & 9/20/2005 & 53 & Florida State & 8/20/2005 \\
4 & Yale & 8/26/2005 & 54 & GWU & 8/27/2005 \\
5 & Cornell & 8/19/2005 & 55 & Johns Hopkins & 8/24/2005 \\
6 & Dartmouth & 9/14/2005 & 56 & Syracuse & 8/24/2005 \\
7 & UPenn & 9/1/2005 & 57 & Notre Dame & 8/19/2005 \\
8 & MIT & 8/28/2005 & 58 & Maryland & 8/24/2005 \\
9 & NYU & 8/28/2005 & 59 & Maine & 9/2/2005 \\
10 & BU & 8/30/2005 & 60 & Smith & 9/2/2005 \\
\hline
11 & Brown & 9/3/2005 & 61 & UC Irvine & 9/19/2005 \\
12 & Princeton & 9/7/2005 & 62 & Villanova & 8/20/2005 \\
13 & UC Berkeley & 8/23/2005 & 63 & Virginia Tech & 8/19/2005 \\
14 & Duke & 8/24/2005 & 64 & UC Riverside & 9/22/2005 \\
15 & Georgetown & 8/27/2005 & 65 & Cal Poly & 9/12/2005 \\
16 & UVA & 8/20/2005 & 66 & Mississippi & 8/17/2005 \\
17 & BC & 8/30/2005 & 67 & Michigan Tech & 8/21/2005 \\
18 & Tufts & 8/31/2005 & 68 & UCSC & 9/17/2005 \\
19 & Northeastern & 9/1/2005 & 69 & Indiana & 8/24/2005 \\
20 & Illinois & 8/18/2005 & 70 & Vermont & 8/26/2005 \\
\hline
21 & Florida & 8/17/2005 & 71 & Auburn & 8/10/2005 \\
22 & Wellesley & 8/29/2005 & 72 & U San Fran & 8/22/2005 \\
23 & Michigan & 8/30/2005 & 73 & Wake Forest & 8/18/2005 \\
24 & Michigan State & 8/25/2005 & 74 & Santa Clara & 9/17/2005 \\
25 & Northwestern & 9/13/2005 & 75 & American & 8/21/2005 \\
26 & UCLA & 9/26/2005 & 76 & Haverford & 8/24/2005 \\
27 & Emory & 8/24/2005 & 77 & William \& Mary & 8/19/2005 \\
28 & UNC & 8/27/2005 & 78 & Miami & 8/17/2005 \\
29 & Tulane & 8/26/2005 & 79 & James Madison & 8/24/2005 \\
30 & UChicago & 9/17/2005 & 80 & UT Austin & 8/26/2005 \\
\hline
31 & Rice & 8/14/2005 & 81 & Simmons & 9/3/2005 \\
32 & WashU & 8/11/2005 & 82 & Binghamton & 8/25/2005 \\
33 & UC Davis & 9/26/2005 & 83 & Temple & 8/22/2005 \\
34 & UC San Diego & 9/15/2005 & 84 & Texas A\&M & 8/22/2005 \\
35 & USC & 8/15/2005 & 85 & Vassar & 8/30/2005 \\
36 & Caltech & 9/18/2005 & 86 & Pepperdine & 8/23/2005 \\
37 & UC Santa Barbara & 9/17/2005 & 87 & Wisconsin & 8/25/2005 \\
38 & Rochester & 8/24/2005 & 88 & Colgate & 8/20/2005 \\
39 & Bucknell & 8/17/2005 & 89 & Rutgers & 8/25/2005 \\
40 & Williams & 8/31/2005 & 90 & Howard & 8/20/2005 \\
\hline
41 & Amherst & 8/28/2005 & 91 & UConn & 8/26/2005 \\
42 & Swarthmore & 8/23/2005 & 92 & UMass & 8/29/2005 \\
43 & Wesleyan & 8/31/2005 & 93 & Baylor & 8/18/2005 \\
44 & Oberlin & 8/30/2005 & 94 & Penn State & 8/25/2005 \\
45 & Middlebury & 9/7/2005 & 95 & Tennessee & 8/13/2005 \\
46 & Hamilton & 8/20/2005 & 96 & Lehigh & 8/25/2005 \\
47 & Bowdoin & 8/27/2005 & 97 & Oklahoma & 8/18/2005 \\
48 & Vanderbilt & 8/20/2005 & 98 & Reed & 8/30/2005 \\
49 & Carnegie Mellon & 8/22/2005 & 99 & Brandeis & 8/28/2005 \\
50 & Georgia & 8/15/2005 & 100 & Trinity & 9/1/2005 \\
\hline
\end{tabular}
\caption{The start of 2005 freshman orientation for the schools in the facebook100 dataset. 
Dates were amassed from individual academic calendars, and 
reflect the start of freshman orientation for non-international students. If such a date could not be found,
dates reflect the day dormitories opened. Failing that, the date was set at 1 week before the start of classes. Summer
pre-orientations were not considered. Calendars from 2005 were found for 71 of the 100 schools. For the remaining schools
a judgement was performed based on more recent calendars and the relative position of orientation/dorms opening to 
Labor Day on the oldest available calendar vs. Labor Day in 2005. All 100 colleges are located in the United States. All sources 
and methodological details are available at \texttt{http://azjacobs.com/fb100/}.}
\label{appxtab2}
\end{table*}

\end{appendix}

\end{document}